\documentstyle[aps,prb,epsf]{revtex}
\begin{document}
\draft 

\title{Quasiclassical Approach to Transport in 
the Vortex State and the Hall Effect. }
\author{A. Houghton and I. Vekhter\cite{me}}
\address{Department of Physics, Brown University, Providence, RI 02912-1843}
\date{\today}
\maketitle

\begin{abstract}
We derive generalized quasiclassical transport equations which 
include the terms 
responsible for the Hall Effect in the vortex state of a clean type-II
superconductor, and calculate the conductivity tensor for an
s-wave superconductor in the high-field
regime. We find that
 below the superconducting transition  the contribution 
to the transverse conductivity due to dynamical 
fluctuations of the order parameter
is compensated by the modification of the quasiparticle contribution.
In this regime the nonlinear behaviour of the Hall angle is governed by the 
change in
the effective quasiparticle scattering rate 
due to the reduction in the density of states at the Fermi 
level.
The connection with experimental results is discussed.

\end{abstract}


\section{INTRODUCTION}
\label{sec:intro}

In the recent years a significant body of work has been devoted
to the better understanding of the Hall effect in the mixed state of type-II 
superconductors, which
has remained a theoretical puzzle 
for almost thirty years \cite{Dorsey,Vinokur}
The phenomenological \cite{BS,NV}  
theories predict that the Hall angle in the flux-flow regime is
either identical to that in the normal state \cite{BS} or constant \cite{NV},
and the underlying 
microscopic basis for recent generalizations \cite{Vinokur}
is not well understood.
Theories which make use of the time-dependent Ginzburg-Landau equations (TDGL)
find that the Hall conductivity is not modified
 in the superconducting state \cite{Dorsey}.
These predictions are at variance with the
strongly nonlinear behaviour (as a function of magnetic field) 
found in experiments performed on both low-$T_c$ materials
\cite{Hagen1,fiory} and the high-$T_c$ cuprates \cite{Hagen1,Ong}.
For dirty superconductors ($l \ll \xi_0$, where $l$ is the mean free path and
$\xi_0$ is the superconducting coherence length),
transport coefficients can be determined from microscopic
theory by a straightforward expansion in powers of the order
parameter, $\Delta$.
The results of such a calculation for the transverse resistivity
\cite{fuku,Ebis}
explain qualitatively the sharp increase in the Hall angle 
below the transition observed in 
experiment (although, to our knowledge, no systematic comparison has
been made), 
and provide the physical basis for a generalized TDGL approach, in which
the relaxation rate is assumed to be complex, rather than purely real,
to allow for a modification of the transverse transport coefficients
\cite{Dorsey,makihall,kopnin}. 
The small parameter in the expansion of the
microscopic equations is
proportional to both the order parameter and the mean free path, 
therefore, it is not small 
in the clean  ($l \gg \xi_0$) 
limit. In this regime a straightforward expansion is not possible;
the TDGL equations are not applicable \cite{makicyrot,HM}, and so an
alternative approach is needed to determine the transverse
transport coefficients. 

In this work we develop an approach to calculate 
the transport coefficients, including the Hall Effect, of
clean type-II superconductors in the vortex state and present the
results of a calculation of the Hall conductivity of a clean 
s-wave superconductor in the mixed state
near the upper critical field, $H_{c2}$.
The
method is based on the quasiclassical approximation
to the microscopic theory, due  originally, 
in the context of superconductivity,
to Eilenberger \cite{E}
and Larkin and Ovchinnikov \cite{LO}, which we
generalize to include the 
terms responsible for the Hall Effect in a charged superfluid.
We solve the equations of this quasiclassical theory to obtain
the longitudinal and transverse resistivities in the mixed state.
We choose to consider an s-wave superconductor
as both the normal state and superconducting properties of the
low-$T_c$ compounds are well known, and comparison between 
theory and experiment is fraught with less ambiguity; however the
approach developed here can easily be generalized to 
consider superconductors with other than 
s-wave symmetry.

The microscopic Green's function contains all the information about 
the single particle properties of the system. In particular, 
it oscillates on length scales of order of the inverse
Fermi wave vector $k_f^{-1}$.
However, when calculating transport coefficients, we are 
for the most part only  interested in the long-wavelength response.
It is then sufficient to determine the envelope  of the 
Green's function rather than its detailed form.
In the quasiclassical approach the rapid
oscillations associated with the presence of the Fermi surface
are integrated out of the basic equations and slower varying quantities
such as external fields or the self energy
are expanded around their values at the Fermi surface.  The resulting
transport like equations contain the microscopic physics relevant
to the problem and are easier to solve. The basic premise of 
quasiclassical transport theory is that all macroscopic physical
quantities vary slowly on a microscopic length scale, and that
all the relevant momenta are small compared to the Fermi
momentum $p_f$. This approximation
has been applied successfully 
to study transport
phenomena in superfluids \cite{Rainer1} and superconductors \cite{Ov}
and to investigate the behavior 
of the unconventional superconductors
\cite{Sauls}. Recently it has been used to analyse 
the most relevant contributions to the Hall effect in a dirty 
superconductor in the limit of
isolated vortices \cite{LO2} as well as to investigate the forces acting on
a single vortex in the clean regime \cite{kopninhall}. 

In the next two sections we present a derivation of the 
generalized quasiclassical equations, which include
 all the terms contributing to
the Hall Effect in the mixed state of a clean type-II superconductor
in the high-field regime.
Section \ref{chap:normal} introduces a general quasiclassical formalism
and the basic ideas involved in the analysis of transverse transport in the
quasiclassical approximation, 
illustrated by application to the simple case of a normal metal. 
We show how the standard Drude results for longitudinal and transverse 
conductivity are obtained within this quasiclassical approximation. 
In Section \ref{chap:eqns} we 
use the same approach to derive a generalization of the
standard quasiclassical approximation for superconductors to include the 
terms responsible for the transverse conductivity and obtain linearized 
transport like equations for a clean superconductor. To solve this 
system of equations near the upper critical field 
we employ the approximation of Brandt, Pesch and Tewordt\cite{BPT} 
(BPT), in which the
normal part of the matrix propagator is replaced by its spatial average
over a unit cell of the vortex lattice, while the exact spatial dependence
of the order parameter is retained. 
Using an operator formalism, we are able to 
solve the leading order equations for the distribution function in
Section \ref{chap:dos}, and obtain the longitudinal and transverse
conductivities within linear response theory in Section \ref{chap:sxx}
and Section \ref{chap:sxy} respectively. In the last section we summarize 
the results and compare them with the existing experimental data.

\section{QUASICLASSICAL APPROACH TO TRANSPORT IN A NORMAL METAL}
\label{chap:normal}

\subsection{Mixed Representation and the Standard Quasiclassical Equations}
\label{subsec:metal1}

Our starting point is the microscopic Dyson's equation
\begin{equation}
\qquad
\biggl[- {\partial \over \partial \tau }-
       \zeta(- i \nabla_{\bf x} )-
\int d^4 y \Sigma (x,y) \biggr]G (y, x')
=\delta (x-x')
\label{eqmetal}
\end{equation}
for the Green's function
\begin{equation}
\label{gf}
G(x, x')=
   -\langle T_{\tau} \psi(x) \psi^{\dagger}(x')\rangle.
\end{equation}
Here $\psi(x)$ and $ \psi^{\dagger}(x)$ are field creation and annihilation 
operators, which depend on the four vector $x=({\bf x},\tau)$, 
angular brackets
denote the statistical average, and the 
operator $T_{\tau}$ arranges the
field operators in ascending order of imaginary time $\tau$.
In Eqn. (\ref{eqmetal}) $\zeta$ is the single particle energy operator, and
$\Sigma$ is the self energy which may be due to interactions
 or impurity scattering, its exact form has to be determined 
from microscopic considerations.
Dyson's equation can also be written in the form
\begin{equation}
\qquad
G (x, x')
\biggl[ {\partial \over \partial \tau '} - \zeta(+i \nabla_{\bf x'}) \biggr]
- \int d^4 y \widehat G (x, y)\widehat\Sigma (y,x')
=\delta (x-x'),
\label{eqmetal1}
\end{equation}
The operators in this equation  are understood to act on the
Green's function on their left.  It should be emphasized that Eqs.
(\ref{eqmetal}) and (\ref{eqmetal1}) contain the same physical information 
and only
differ in the form of writing, i.e. the same function $G$
satisfies both. We will use the terms 
right-hand and left-hand Dyson's equation for Eqs. (\ref{eqmetal})
and (\ref{eqmetal1}) respectively. 

 The derivation of the quasiclassical equations
given here follows the general approach
of Rainer and Serene \cite{Rainer1,Rainer2} and 
Eckern and Schmid \cite{Schmid}.
First we consider the linear response of a 
metal to a constant uniform electric field 
described by a vector potential
${\bf A}(\tau)={\bf A}\exp(i \omega_0 \tau)$. 
To incorporate the vector potential
into the microscopic equations 
we replace the momentum operator by its gauge invariant counterpart
$\zeta(-i \nabla_{\bf x}) \rightarrow 
\zeta(-i \nabla_{\bf x}- e{\bf {A}}(\tau))$,
and expand this expression to obtain terms linear in the
external field. To integrate out the rapid
oscillations associated with the presence of the Fermi surface
we 
first
change variables from ${\bf x}$ and ${\bf x'}$ to center of mass and
relative coordinates ${\bf R}=({\bf x}+{\bf x'})/2$ and 
 ${\bf r}={\bf x}-{\bf x'}$,
and carry out a Fourier transformation in the latter according to
\begin{equation}
\label{mix1}
G({\bf p},{\bf R}) = \int d^3 {\bf r} G({\bf R}+{{\bf r} \over 2},
                                 {\bf R} - {{\bf r} \over 2}) 
                                   \exp(-i{\bf p}{\bf r})
\end{equation}
In a translationally invariant system the Green's function only depends  on 
the relative coordinate. Therefore, dependence on the  position of the 
center of
mass ${\bf R}$ appears only
in the presence of external fields. To treat the effect of slowly
varying fields quasiclassically we expand in quantities varying on 
the length scale of the wavelength of these fields, which is equivalent to
expanding in powers of $\nabla_{\bf R}$. 
If  $A({\bf x}, -i \nabla_{\bf x})$ is  a local operator which
depends only on position and momentum and  acts on the Green's function
$G(x, x')$, then
\begin{eqnarray}
&&\int d^3 {\bf r}\exp(-i{\bf p}{\bf r})
A({\bf x}, -i \nabla_{\bf x})G(x, x')
\label{transform}
= \int d^3 {\bf r}  \exp(-i{\bf p}{\bf r})
A\bigl({\bf R} + {{\bf r}\over 2}, -i \nabla_{\bf r} 
-{i \over 2}\nabla_{\bf R}\bigr)
G \bigl({\bf R} + {{\bf r}\over 2},{\bf R} - {{\bf r}\over 2}\bigr)
\\
\nonumber
&&\qquad= \int d^3 {\bf r}
A\bigl({\bf R} + {i\over 2}\nabla_{\bf p},
{\bf p}-{i \over 2}\nabla_{\bf R}\bigr)
G \bigl({\bf R} + {{\bf r}\over 2},{\bf R} - {{\bf r}\over 2}\bigr)
 \exp(-i{\bf p}{\bf r})
=A\bigl({\bf R} + {i\over 2}\nabla_{\bf p},
{\bf p}-{i \over 2}\nabla_{\bf R}\bigr)
G({\bf p},{\bf R}).
\end{eqnarray}
The final expression can be written as $A\circ G$,
where the ``circle-product'' is defined as \cite{Rainer2,Schmid}
\begin{equation}
\label{oprod}
A({\bf p},{\bf R}) \circ B({\bf p},{\bf R})
=    \exp ({i \over 2}( \nabla_{{\bf p}_2} \nabla_{{\bf R}_1} -
                                 \nabla_{{\bf p}_1} \nabla_{{\bf R}_2}))
          A({\bf p}_1,{\bf R}_1)B({\bf p}_2,{\bf R}_2)
                                 |_{{\bf R}_1={\bf R}_2={\bf R}},
\end{equation}
Using this definition the right and left-hand Dyson's equations
can be written in terms of the mixed set of variables ${\bf p}$ and ${\bf R}$ 
as
\begin{eqnarray}
\label{eqmix1}
&&\biggl[- {\partial \over \partial \tau }-
       \zeta({\bf p}- e{\bf {A}}(\tau))\biggr] \circ 
	G({\bf p},{\bf R};\tau, \tau')
- \int d\tau_1 \Sigma ({\bf p},{\bf R};\tau, \tau_1) \circ
	 	G({\bf p},{\bf R};\tau_1, \tau')=
   \delta(\tau-\tau')
\\
\label{eqmix2}
&&G({\bf p},{\bf R};\tau, \tau') \circ 
\biggl[ {\partial \over \partial  \tau '} - 
\zeta({\bf p}- e{\bf {A}}(\tau'))\biggr]
-\int d\tau_1 G({\bf p},{\bf R};\tau, \tau_1) \circ
 \Sigma ({\bf p},{\bf R};\tau_1, \tau')=
   \delta(\tau-\tau').
\end{eqnarray}
Direct expansion of Eqs. (\ref{eqmix1}) and (\ref{eqmix2}) 
in powers  of the spatial gradient 
is not possible since in the definition Eq. (\ref{oprod}) 
of the circle-product this gradient is coupled to derivatives with respect
to momentum, and 
the Green's function varies rapidly with momentum near $p_f$. 
To avoid this  problem we make a transformation from the set of
variables $({\bf p},{\bf R})$ to the set $(s,\zeta,{\bf R})$, where
$s$ is a parameterization of the Fermi surface, 
and integrate the quantum mechanical equations 
over the quasiparticle energy $\zeta$ before expanding. 
The  integrated Green's
function 
\begin{equation}
\label{g}
g(s,{\bf R};\tau, \tau')=
{1 \over \pi} \int d\zeta
 G({\bf p}, {\bf R};\tau, \tau')
\end{equation}
only depends on the 
components of momentum parallel to the Fermi
surface and the remaining dependence on ${\bf p}$ and ${\bf R}$ is slow.
We now transform Eqs. (\ref{eqmix1}) 
and (\ref{eqmix2}) for the full microscopic Green's function $G$ 
into equations for the quasiclassical propagator $g$. This quasiclassical 
propagator will play the role of a distribution function in the 
resulting transport like equation.

Let us first compare terms of zeroth order in the gradient expansion
of Eqs. (\ref{eqmix1})and (\ref{eqmix2}).
 Since the imaginary time $\tau$ varies
between $0$ and $1/T$, where $T$ is temperature, the first term 
in the equation gives, after integration, a contribution of  order $Tg$. 
 If we assume that the self energy
varies slowly for momenta close to the Fermi momentum 
$|{\bf p}|\approx p_f$,
we can approximate
\begin{equation}
\int d \zeta \Sigma ({\bf p}, \ldots)
		G({\bf p}, \ldots) \approx
 \Sigma ({\bf p}_f, \ldots)\int d \zeta G({\bf p}, \ldots)
\equiv\sigma(s, \ldots) g(s, \ldots).
\end{equation}
On the other hand, the term involving $\zeta$ gives a much larger contribution
since the integration region includes $\zeta \sim \epsilon_f$. Because
of this term and the delta function on the right-hand side 
the equations cannot be integrated directly.
Instead, 
we subtract Eq. (\ref{eqmix2}) 
from
Eq. (\ref{eqmix1}) to obtain a homogeneous form before integrating
term by term and expanding in the gradients.
The zeroth order term involving $\zeta G$ then cancels. 
Expanding  to 
first order we obtain
\begin{equation}
\label{expand1}
\int d \zeta \bigl[ G \circ
\zeta({\bf p}- e{\bf {A}}(\tau'))
- \zeta({\bf p}- e{\bf {A}}(\tau))\circ G \bigr]
\approx
- e {\bf v} \biggl[{\bf {A}}(\tau')-{\bf {A}}(\tau)\biggr]g
+ i {\bf v} \nabla_{\bf R} g,
\end{equation}
where the Fermi velocity is defined as
\begin{equation} 
\label{vf}
{\bf v}= {\partial \zeta \over \partial {\bf p}} ({\bf p}_f).
\end{equation}
If the spatial dependence of the distribution function is 
determined by the wave vector ${\bf q}$ of an external field, the
product ${\bf v}{\bf q}$ is not necessarily small compared
to the temperature and the self energy, so that this term has to be
retained in the leading order equation. 
Since the small parameters in the expansion are of order $1/(k_f \lambda)$,
where $\lambda$ is a typical wavelength of the electric field, for the terms
involving the external vector potential,
or, if the self energy is due to impurity scattering,
$1/(k_f l)$,
$\zeta$ must always be expanded to one order higher
in small quantities
than other terms in order to  obtain a contribution of similar order.
It should also be emphasized that, since there are several small
parameters in the problem, it may be necessary to expand
terms to different order in gradients to account for 
all  the contributions to a particular physical quantity.

It is convenient to  Fourier decompose
the integrated Green's
function into  Matsubara frequencies 
\begin{equation} 
g(s,{\bf R};\tau, \tau')=
T \sum_{n,n'}g(s,{\bf R};\omega_n, \omega_{n'})
\exp(-i\omega_n\tau+i\omega_{n'}\tau'),
\end{equation}
where 
$\omega_n$ are the fermionic
frequencies $\omega_n = (2n +1)\pi T$.
Then  the basic transport equation of the quasiclassical
formalism becomes
\begin{eqnarray}
\nonumber
&&\biggl[i\omega_n
- i\omega_{n'}  +i {\bf v} (s) \nabla_{\bf R}\biggr]
g(s,{\bf R};\omega_n,\omega_{n'})
+e {\bf v} {\bf {A}}
\biggl[g(s,{\bf R};\omega_n-\omega_0,\omega_{n'})
          -g(s,{\bf R};\omega_n,\omega_{n'}+\omega_0)\biggr]
\\
\label{basicmetal}
&&-T\sum_{\omega_k}
     \biggl[\sigma(s,{\bf R};\omega_n,\omega_k)
 g(s,{\bf R};\omega_k,\omega_{n'})
- g(s,{\bf R};\omega_n,\omega_k)
\sigma(s,{\bf R};\omega_k,\omega_{n'})\biggr]
=0.
\end{eqnarray}
The exact form of the self energy, $\sigma$, is determined from
microscopic theory.  In principle, all higher order terms
in the spatial gradient can be included in this equation consistently
using the definition of the circle product.

It should be noted
that, in the absence of a perturbing potential, 
 or impurity scattering leading to the appearance of
the self energy, 
the Green's function is
independent of the coordinate ${\bf R}$ and 
is diagonal in frequency space, and therefore
Eq. (\ref{basicmetal}) is trivially satisfied by any function
$g$. This is not surprising since in subtracting the
right hand  Dyson's equation from the left hand equation
the information 
about a particular
solution of the inhomogeneous equation has been lost. The particular 
solution describes the unperturbed non-interacting electron gas,
and is obtained by integrating the function
$G_0=[i \omega_n - \zeta ({\bf p})]^{-1}$
over the quasiparticle energy
to find
the quasiclassical distribution function of a normal metal, 
$g_0= - i {\rm sgn}(\omega_n)$. 
This function serves as input for any perturbative
approach to transport in a metal.

\subsection{Semiclassical Treatment of the Magnetic Field and  the Lorentz
Force.}
\label{sec:lorentz}

Eq. (\ref{basicmetal}) is sufficient to analyse longitudinal 
transport in a normal metal but it has to be generalized to determine 
 the 
Hall conductivity. If the vector potential ${\cal A}({\bf R})$ describing the
 magnetic
field is taken to be of order ``{\it small}'', the field itself,
${\bf H}=\nabla \times {\cal A}$, becomes of order ``$(small)^2$'' 
and the Lorentz force, which is proportional
to both electric and magnetic fields, 
disappears from the perturbative expansion of the quasiclassical equations. 
This observation led Rainer \cite{Rainer2} to point out that 
in order to analyse the Hall Effect
in a normal metal, the vector potential 
${\cal A}({\bf R})$  must
 be considered as a leading order quantity and should be included in 
the equations 
semiclassically
rather than being treated perturbatively.
 Now the quasiparticle energy $\zeta$
depends on the generalized momentum ${\bf p}- e {\bf A}-e {\cal A}({\bf R})$.
This replacement is exact. The semiclassical 
approximation, which is applicable in the long wavelength limit where 
the quasiclassical approach 
is appropriate,
treats the momentum operator as a $c$-number.
Therefore in the transformations described in 
Eq. (\ref{transform}) the momentum ${\bf p}$ and 
the coordinate ${\bf R}$ are
no longer independent variables, rather, they are coupled by the 
presence of  the vector potential,
which depends upon the coordinates. As a result the gradient
expansion of the integrated Green's function cannot be carried out
 independently
in the Fermi surface parameterization $s$ and the spatial variable ${\bf R}$.
For a general transformation of variables from the set $({\bf p}, {\bf R})$ 
to  the set $(\zeta,s_i, {\bf R})$
\begin{eqnarray}
\label{partialR}
{\partial \over \partial R_\alpha} &=&
 {\partial \over \partial R_\alpha} + {\partial \zeta \over \partial R_\alpha}
                                             {\partial \over \partial \zeta}
     + {\partial s_i \over \partial R_\alpha}{\partial \over \partial s_i}
\\
{\partial \over \partial  p_\alpha}&=&
     {\partial \zeta \over \partial  p_\alpha}{\partial \over \partial \zeta}
   + {\partial s_i \over \partial p_\alpha}{\partial \over \partial s_i}
\end{eqnarray}
where the derivatives on the right hand side are computed at constant $\zeta,
s, {\bf R}$ rather than ${\bf p}, {\bf R}$.
Using the explicit semiclassical ${\bf R}$-dependence of $\zeta$ and $s$
\begin{eqnarray}
{\partial \zeta\over \partial R_\alpha} &=&
-e{\partial \zeta\over \partial p_\beta}
{\partial {\cal A}_\beta\over \partial R_\alpha}=
-ev_\beta {\partial {\cal A}_\beta\over \partial R_\alpha},
\\
{\partial s_i\over \partial R_\alpha} &=&
-e{\partial s_i\over \partial p_\beta}
{\partial {\cal A}_\beta\over \partial R_\alpha},
\end{eqnarray}
we obtain from the expansion 
of the terms involving $\zeta$ 
\begin{eqnarray}
\label{lorentz}
&&\int d\zeta \biggl[
-\zeta({\bf p}-e{\bf A}(\tau) - e{\cal A}({\bf R}))
\circ G + G \circ 
\zeta({\bf p}-e{\bf A}(\tau') - e{\cal A}({\bf R}))\biggr]
\\ 
\nonumber
&&\approx
\int d\zeta \biggl(
-\zeta({\bf p}-e{\bf A}(\tau) - e{\cal A}({\bf R})) G + 
\zeta({\bf p}-e{\bf A}(\tau') - e{\cal A}({\bf R}))G
+i
\biggl[{\partial\zeta\over\partial {\bf p}}
	{\partial G \over \partial {\bf R}}
- {\partial\zeta\over\partial {\bf R}}
 {\partial G \over \partial{\bf p}}\biggr]\biggr)
\\ 
\nonumber
&&\rightarrow
- e {\bf v} \biggl[{\bf {A}}(\tau')-{\bf {A}}(\tau)\biggr]g
+ i {\bf v} \nabla_{\bf R} g
+ i e ( {\bf v} \times {\bf H} ) {\partial g \over \partial {\bf p}_{\|}}, 
\end{eqnarray}
where ${\bf p}_{\|}$ denotes the component of the momentum ${\bf p}$ parallel 
to the Fermi surface. In the last line of Eq.(\ref{lorentz}) we have used 
the result
\begin{eqnarray}
\nonumber
\biggl[{\partial\zeta\over\partial {\bf p}}
	{\partial G \over \partial {\bf R}}
- {\partial\zeta\over\partial {\bf R}}
 {\partial G \over \partial{\bf p}}\biggr]&=&
{\partial\zeta\over\partial p_\alpha}{\partial G \over \partial R_\alpha}
+\biggl[{\partial\zeta\over\partial p_\alpha}
{\partial s_i\over \partial R_\alpha}-
{\partial\zeta\over\partial R_\alpha}
{\partial s_i\over \partial p_\alpha}\biggr]
{\partial G \over \partial s_i}
\\
&=&
v_\alpha {\partial G \over \partial R_\alpha}
+ev_\alpha \biggl[{\partial {\cal A}_\alpha\over\partial R_\beta}-
{\partial {\cal A}_\beta\over\partial R_\alpha}\biggr]
{\partial s_i\over \partial p_\beta}{\partial G \over \partial s_i}
={\bf v} \nabla_{\bf R} G +
e ( {\bf v} \times {\bf H} ) {\partial G \over \partial {\bf p}_{\|}}.
\end{eqnarray}
The new term is the familiar Lorentz force driving 
term of the classical
Boltzmann
transport equation. 
Here it appears from taking into account correctly the
semiclassical dependence of the momentum on the external field. 
The  basic quasiclassical equation Eqn. (\ref{basicmetal}) 
now takes the form
\begin{eqnarray}
\nonumber
&&\biggl[i{\bf v} (s) \nabla_{\bf R}+
i e({\bf v} \times {\bf H} ){\partial \over \partial {\bf p}_{\|}}\biggr]
g(s,{\bf R};\omega_n,\omega_{n'})+
\biggl[ i\omega_n
- i\omega_{n'}\biggr]g(s,{\bf R};\omega_n,\omega_{n'})
\\
\label{nbase1}
&&-T\sum_{\omega_k}
     \biggl[\sigma(s,{\bf R};\omega_n,\omega_k)
 g(s,{\bf R};\omega_k,\omega_{n'})
- g(s,{\bf R};\omega_n,\omega_k)
\sigma(s,{\bf R};\omega_k,\omega_{n'})\biggr]
\\
\nonumber
&&+e {\bf v} {\bf {A}}
\biggl[g(s,{\bf R};\omega_n-\omega_0,\omega_{n'})
          -g(s,{\bf R};\omega_n,\omega_{n'}+\omega_0)\biggr]
=0.
\end{eqnarray}

\subsection{Linear Response}
\label{sec:linmetal}

In general Eq. (\ref{nbase1}) is a nonlinear equation. 
To calculate transport coefficients it is sufficient to
keep only the terms linear in the external perturbation --
in this case in the electric field --  and determine the
Green's function $g$ within  linear response.
We separate the propagator 
into a leading term and a part linear in the
perturbing potential 
\begin{equation}
\label{glinear}
g=g_0(\omega_n) 
   {1 \over T} \delta_{\omega_n, \omega_{n'}}+
   g^{(1)}(s, {\bf R}; \omega_n, \omega_0)
   {1 \over T} \delta_{\omega_n, \omega_{n'} + \omega_0}. 
\end{equation}
If the self energy is due to elastic impurity scattering, it can be 
separated in a similar way into $\sigma_0$ and
$\sigma^{(1)}$.
As noted the equation for the leading order 
terms $g_0$ and $\sigma_0$ is 
satisfied trivially; the terms of linear order are given by
\begin{equation}
\label{nlin}
\biggl[
i \omega_0 + \sigma_0 (-) -  \sigma_0 
+i {\bf v} (s)  \nabla_{\bf R}\biggr]
g^{(1)}
+i e ( {\bf v} \times {\bf H} ) {\partial g^{(1)} \over \partial {\bf p}_{\|}}
=\bigl(e {\bf v}{\bf A}+ \sigma^{(1)} \bigr)
\bigl(g_0(-) - g_0\bigr),
\end{equation}
here we have used a shorthand notation
$ g_0=g_0(\omega_n)$ and $g_0(-)=g_0(\omega_n-\omega_0)$.
This equation is the basis for the analysis of transport in a
normal metal. It has to be solved together with the self consistency
condition relating the change in the self energy to the
modification of the Green's
function $ g^{(1)}$. 

Since, throughout this work, we will be concerned with the  electrical 
conductivity, we have
to define the current in terms of the distribution function.
It is well known \cite{AGD} that, if in the
microscopic equation for the current density
\begin{equation}
\label{qmcurr}
{\bf j}({\bf x})={ e \over m} T \sum_{\omega_n} \int d^3 {\bf p} {\bf p}
G^{(1)}({\bf p}, {\bf R}\rightarrow{\bf x};\omega_n)
- {N e^2\over m}{\bf A}
\end{equation}
the integration over energy is carried out before summing over
frequencies, the contribution from the
high energy regions (far above and below the Fermi surface)
exactly cancels the diamagnetic term in
Eq. (\ref{qmcurr}). Then the quasiclassical
expression for  the current becomes \cite{AGD,Eliash}
\begin{equation}
\label{current}
{\bf j}({\bf R})=
\pi N(0) e T\sum_{\omega} \int d^2 s {\bf v} (s) g^{(1)}(s, {\bf R};\omega),
\end{equation}
where $N(0)$ is the density of states at the Fermi surface.
The problem of calculating the transport coefficients of a 
normal metal is now fully defined.

\subsection{Conductivity of a Normal Metal}
\label{sec:normalsigma}

As an example of the usefulness of the quasiclassical method  we
will use it to determine the conductivity tensor of a normal metal
in a magnetic
field.
We consider an experimental arrangement with constant electric and magnetic
fields ${\bf E}=E \widehat{\bf x}$ and ${\bf H}=H \widehat {\bf z}$.
We also assume a spherical Fermi surface  
\begin{equation}
{\bf v}=v(\sin\theta\cos\phi,\sin\theta\sin\phi,\cos\theta)
\end{equation}
and include the effect  of
isotropic impurity scattering in the Born approximation, so that
the self energy is given by
\begin{equation}
\label{tau}
\sigma = {1 \over 2\tau} \int d^2 s g
\end{equation}
where $\tau$ is the quasiparticle lifetime. 
The unperturbed Green's function is given by
$g_0=-i {\rm sgn}(\omega_n)$, and therefore
$\sigma_0= -i{\rm sgn}(\omega_n) / 2 \tau$.

First consider the longitudinal dc conductivity. In the absence of a
magnetic field  Eq. (\ref{nlin}), becomes
\begin{equation}
\label{eqnormal}
\Bigl[
i \omega_0 + {i \over 2 \tau} ({\rm sgn}(\omega_n) -
                                {\rm sgn}(\omega_n - \omega_0) )
\Bigr] g^{(1)}=
-i \bigl(e {\bf vA}+\sigma^{(1)}\bigr) ({\rm sgn}(\omega_n) -
                                {\rm sgn}(\omega_n - \omega_0) ).
\end{equation}
Since the driving term in Eq. (\ref{eqnormal}) is proportional
to ${\bf vA}$, it is evident that the angular dependence of 
$g^{(1)}$ is given by that dot product,
and  there is no correction to the self energy since
the angular average of $ g^{(1)}$ vanishes. 
Then it is obvious from Eq. (\ref{eqnormal})
that $g^{(1)}=0$ when $\omega_n$ and 
$\omega_n - \omega_0$ have the same
sign. Otherwise, in the intermediate 
frequency region where $\omega_0>\omega_n>0$,
\begin{equation} 
\label{nlong} 
g^{(1)}= -
{2 e {\bf v} (s) {\bf A} \over {\omega_0 +{1 / \tau}}}.
\end{equation}
Integrating over the Fermi surface, 
carrying out the summation in the definition of
current density, 
and analytically continuing
to the real 
external frequency
according to $i\omega_0\rightarrow \bar\omega +i\delta$, 
in the dc-limit
($\bar\omega\rightarrow 0$) 
we recover from this  solution
the standard Drude theory result for the current
\begin{equation}
{\bf j}={1 \over 3 } N(0) e^2 v^2 \tau {\bf E}=\sigma_n {\bf E}.
\end{equation}

We now turn on the magnetic field. Writing the expression for the
Lorentz force in spherical coordinates 
it is easy to check that
\begin{equation}
\label{lorenz}
e({\bf v} \times {\bf H} ){\partial  \over \partial {\bf p}_{\|}}=
-\omega_c {\partial  \over \partial\phi},
\end{equation}
and the linearized transport equation becomes
\begin{equation}
\bigl[
i \omega_0 + {i \over 2 \tau} ({\rm sgn}(\omega_n) -
                                {\rm sgn}(\omega_n - \omega_0) )
- i \omega_c {\partial\over \partial \phi}\bigr] g^{(1)}
=
-i e {\bf v}(s) {\bf A} ({\rm sgn}(\omega_n) -
                                {\rm sgn}(\omega_n - \omega_0) ).
\label{nhall}
\end{equation}
Again, the response function is non-zero in the intermediate region
only. 
In the regime $\omega_c \tau \ll 1$ it is sufficient to solve the equation 
perturbatively, namely,
\begin{eqnarray}
\label{g1h}
g^{(1)}_H&=& -{2 e {\bf v} (s) {\bf A} \over {\omega_0 +{1 / \tau}}}
           + \delta g
\\
\delta g &=&
      {2 e^2  \over ({\omega_0 +{1 / \tau}})^2}
       \omega_c v A\sin\theta\sin\phi.
\end{eqnarray}
The transverse current obtained from the correction $\delta g$ is, as
expected,
\begin{equation}
 j_y = -\sigma_n \omega_c \tau E.
\end{equation}
We have therefore reproduced the results of the Drude theory using the 
quasiclassical formalism. 

\section{QUASICLASSICAL EQUATIONS FOR A SUPERCONDUCTOR}
\label{chap:eqns}

In this section we generalize the approach developed in Section
\ref{chap:normal} to derive
a set of quasiclassical equations which can be used to analyse both 
longitudinal and transverse transport in superconductors. 

\subsection{Gorkov equations}

Gorkov's equations \cite{gorkov1} for a matrix Green's function
$\widehat G$ replace Dyson's equations in a fully microscopic approach
to a superconductor.
The diagonal elements of the matrix Green's function
\begin{equation}
\widehat G=\pmatrix{G& -F \cr F^{\dagger} & \bar G}
\end{equation}
are the particle and hole propagators, 
\begin{equation}
\label{g22}
\bar G(x,x')= G(x',x);
\end{equation}
for singlet pairing, 
the off diagonal elements are related to the 
probability amplitudes
for the
 destruction or 
creation of a Cooper pair by 
\begin{eqnarray}
\label{F}
(i \widehat\sigma_y)_{\alpha\beta} F(x, x')&=&
	 -\langle T_{\tau}\psi_{\alpha}(x)\psi_{\beta}(x')\rangle
\\
\label{Fdag}
(i \widehat\sigma_y)_{\alpha\beta} F^{\dagger}(x, x')&=&
	 \langle T_{\tau}\psi^{\dagger}_{\alpha}(x)
	\psi^{\dagger}_{\beta}(x')\rangle
\end{eqnarray}
where $\widehat\sigma_y$ is the Pauli matrix.
Then the right and left-hand Gorkov equations are
\begin{eqnarray}
\qquad
\biggl[- {\partial \over \partial \tau }\widehat\sigma_z -
       \zeta(- i \nabla_{\bf x} \widehat\sigma_z)
+\widehat\Delta (x) \biggr] \widehat G (x, x')-
\int d^4 y \widehat\Sigma (x,y) \widehat G (y, x')
=\delta (x-x') {\tt 1}
\label{geq}
\\
\qquad
\widehat G (x, x')
\biggl[ {\partial \over \partial \tau '} \widehat\sigma_z - 
\zeta(+i \nabla_{\bf x'}\widehat\sigma_z) +
\widehat \Delta (x') \biggr] 
- \int d^4 y \widehat G (x, y)\widehat\Sigma (y,x')
=\delta (x-x') {\tt 1}.
\label{geq1}
\end{eqnarray}
The matrix order parameter
\begin{equation}
\label{D}
\widehat\Delta=\pmatrix{0&\Delta\cr -\Delta^\star &0\cr}
\end{equation}
is related to the off-diagonal elements of the Green's function by
\begin{eqnarray}
\label{D0}
\Delta (x)& =& {\rm g} F(x+0,x)
\\
\label{DD}
\Delta^{\star}(x)&=& {\rm g} F^{\dagger}(x+0,x),
\end{eqnarray}
where g is the coupling constant.

\subsection{Quasiclassical Approximation}

The general approach to the derivation of 
the quasiclassical equations for superconductors
is exactly the same as that of Section
\ref{chap:normal}.
We introduce the vector potentials ${\bf A}(\tau)$ and ${\cal A} ({\bf x})$
of an electric and magnetic field into the energy operator, transform the
equations to a set of ``mixed'' variables ${\bf p}$ and ${\bf R}$
by performing a Fourier transform in the relative coordinate, and
expand in gradients with respect to the center of mass coordinate,
after integration over the quasiparticle energy.

Expanding the circle product $\zeta(-i\nabla_{\bf x})\circ G$
to first order in gradients, we obtain
\begin{equation}
\label{expandx}
\int d^3{\bf r} \exp (- i{\bf pr})
\zeta(-i \nabla_{\bf x}- {\cal A}({\bf x}) - {\bf A}) G
\rightarrow
\biggl[\zeta({\bf p}) - {i\over 2}{\bf v}
\Bigl(\nabla_{\bf R} - 2ie {\cal A}({\bf R})\Bigr) - {\bf v}{\bf A}
-{ie\over 2} \bigl({\bf v}\times {\bf H}\bigr)
	{\partial \over \partial {\bf p}}\biggr]
G({\bf p},{\bf R}).
\end{equation}
On the other hand  on expanding the operator
$\zeta(+i\nabla_{\bf x'})$  
the combination
${\bf p}+i\nabla_{\bf R}$ rather than
${\bf p}-i\nabla_{\bf R}$ appears after
Fourier transform in ${\bf r}$.  
Consequently the magnetic field dependent terms arising from the
expansion of $\nabla_{\bf R}$ in Eq. (\ref{partialR})
have the opposite sign, and    
\begin{equation}
\label{expandprime}
\int d^3{\bf r} \exp (- i{\bf pr})
\zeta(+i \nabla_{\bf x'}- {\cal A}({\bf x'}) - {\bf A}) G
\rightarrow
\biggl[\zeta({\bf p}) + {i\over 2}{\bf v}
(\nabla_{\bf R} + 2ie {\cal A}({\bf R})) - {\bf v}{\bf A}(\tau')
+{ie\over 2} \bigl[{\bf v}\times {\bf H}\bigr]
	{\partial \over \partial {\bf p}}\biggr]
G({\bf p},{\bf R}).
\end{equation}
Subtracting Eq. (\ref{expandprime}) from equation 
(\ref{expandx}) we regain the result of
the Section \ref{sec:lorentz}.
The vector potential
${\cal A}$ appears in the expansions in different gauge
invariant combinations. This can be easily understood if we
remember that operator $\zeta(-i \nabla_{\bf x})$ acts on
the  annihilation operator $\psi$ while operator
$\zeta(+i \nabla_{\bf x'})$ acts on the creation operator $\psi^{\dagger}$.
Then the time evolution of the operators  describes
the motion of particles and holes respectively, and
the appropriate gauge invariant derivative is different in each case.
To determine the transverse conductivity all  contributions
of the order of the cyclotron frequency 
$\omega_c=eH/mc$ have to be included in the
equations. 
In a type-II superconductor in the vortex state 
the coherence length $\xi_0$ 
sets the length scale for  spatial change of the
order parameter.
Near the upper critical field
$H_{c2}$
 the magnetic length 
$\Lambda = (2eH)^{-1/2} \simeq \xi_0$,
this immediately implies that the expansion of the operator 
$\zeta({\bf p})$
has to be carried out not to first, but 
to second order in spatial derivatives.
The second order derivative of $\zeta$ with respect to momentum is, 
by definition, the inverse effective mass tensor, which in the simple case
of a spherical Fermi surface becomes equal to the inverse effective mass
$m$. In the expansion this term is coupled to square of
the spatial gradient, so that its contribution 
\begin{equation}
{\partial^2 \zeta \over \partial p_\alpha\partial p_\beta}
\nabla_\alpha \nabla_\beta G
\sim
{1\over m\Lambda^2}G \propto \omega_c G
\end{equation}
is comparable to that of the Lorentz force term and has to be
taken into account.
Neglecting  terms
quadratic in the electric and magnetic fields
and assuming a Fermi surface with the reflection symmetry
 $\zeta({\bf p})= \zeta(-{\bf p})$
we obtain the expansion of the quasiparticle energy
operator 
\begin{eqnarray}
\label{zeta1}
&&\zeta(-i \nabla_{\bf x})
\rightarrow
\zeta({\bf p}) - {i\over 2}{\bf v}
\bigl(\nabla - 2ie {\cal A}({\bf R})\bigr) - {\bf v}{\bf A}(\tau)
-{ie\over 2}\bigl[{\bf v}\times {\bf H}\bigr]
	{\partial \over \partial {\bf p}}
\\
\nonumber
&&\qquad\qquad\qquad\qquad
-{1\over 8 m}\bigl(\nabla-2i e {\cal A}({\bf R})\bigr)^2
+{i e \over 2 m}{\bf A}\bigl(\nabla-2i e {\cal A}({\bf R})\bigr)
\\
\label{zeta2}
&&\zeta(+i \nabla_{\bf x})
\rightarrow
\zeta({\bf p}) - {i\over 2}{\bf v}
\bigl(\nabla + 2ie {\cal A}({\bf R})\bigr) + {\bf v}{\bf A}(\tau)
+{ie\over 2} \bigl[{\bf v}\times {\bf H}\bigr]
	{\partial \over \partial {\bf p}}
\\
\nonumber
&&\qquad\qquad\qquad\qquad
-{1\over 8 m}\bigl(\nabla+2i e {\cal A}({\bf R})\bigr)^2
-{i e \over 2 m}{\bf A}\bigl(\nabla+2i e {\cal A}({\bf R})\bigr)
\end{eqnarray}
and similar expressions for the operators
$\zeta(+i \nabla_{\bf x'})$ and $\zeta(-i \nabla_{\bf x'})$.

Now consider the remaining terms in the expansion of
the  microscopic Eqs. (\ref{geq}) and (\ref{geq1}). 
Here we  are concerned with the change 
in the Hall conductivity of a superconductor
relative to the normal state value. This change involves the 
magnitude of the superconducting order parameter $\Delta$,
which appears in our analysis in the 
dimensionless combination 
 $(\Lambda \Delta /v)$. It is then  easily seen  that
 linear terms in the gradient expansion of the order parameter
have to be retained in the equation since a typical term in the expansion
\begin{equation}
{\partial\widehat\Delta\over\partial{\bf R}}
{\partial \widehat G \over\partial{\bf p}}
\propto
{\Delta\over\Lambda}{1\over mv}\widehat G
\approx \omega_c \bigl({\Lambda\Delta\over v}\bigr)\widehat G,
\end{equation}
will contribute significantly 
to the change of transverse conductivity upon entering
the superconducting state. Expanding to  first order in the gradients
we obtain from Eq. (\ref{oprod})
\begin{eqnarray}
\nonumber
\widehat \Delta(x)\widehat G(x,x')- \widehat G(x,x') \widehat \Delta(x')
&\rightarrow&
\widehat \Delta({\bf R},\tau)\widehat G({\bf p}, {\bf R})
-\widehat G({\bf p}, {\bf R})\widehat \Delta({\bf R},\tau')
\\
\label{dexpand}
&&+ {i \over 2}\biggl[ {\partial\widehat\Delta\over\partial {\bf R}}
{\partial \widehat G \over\partial{\bf p}}
+{\partial \widehat G \over\partial {\bf p}}
{\partial\widehat\Delta\over\partial {\bf R}}\biggr]
-{i \over 2}\biggl[ {\partial\widehat\Delta\over\partial {\bf p}}
	{\partial \widehat G \over\partial {\bf R}}
+{\partial \widehat G \over\partial {\bf R}}
 {\partial\widehat\Delta\over\partial {\bf p}}\biggl],
\end{eqnarray}
and, similarly,
\begin{eqnarray}
\nonumber
&&\int d^4 y
\biggl(-\widehat\Sigma (x,y) \widehat G (y, x')+
\widehat G (x, y)\widehat\Sigma (y,x')
\biggr)
\\
\label{sexpand}
&&\rightarrow
\int d\tau_1\biggl(\widehat \Sigma ({\bf p},{\bf R};\tau,\tau_1)
\widehat G({\bf p}, {\bf R};\tau_1,\tau')
-\widehat G({\bf p}, {\bf R};\tau, \tau_1)
\widehat \Sigma({\bf p},{\bf R},\tau_1,\tau')
\\
\nonumber
&&+ {i \over 2}\biggl[ {\partial\widehat\Sigma\over\partial {\bf R}}
{\partial \widehat G \over\partial{\bf p}}
+{\partial \widehat G \over\partial {\bf p}}
{\partial\widehat\Sigma\over\partial {\bf R}}\biggr]
-{i \over 2}\biggl[ {\partial\widehat\Sigma\over\partial {\bf p}}
	{\partial \widehat G \over\partial {\bf R}}
+{\partial \widehat G \over\partial {\bf R}}
 {\partial\widehat\Sigma\over\partial {\bf p}}\biggr]\biggr).
\end{eqnarray}

Using the results of Eqs. (\ref{zeta1})--(\ref{zeta2}),
(\ref{dexpand}) and (\ref{sexpand}),
subtracting the left-hand Gorkov-Dyson equation
from the right-hand equation, and integrating over the quasiparticle
energy, we obtain
the quasiclassical transport equation for a 
superconductor, which can be written 
using the matrix notation as follows
\begin{eqnarray}
\nonumber
&&
i \omega_n \widehat\sigma_z \widehat g - i\omega_{n'} 
\widehat g \widehat\sigma_z
+\widehat\Delta\widehat g - \widehat g \widehat\Delta
+i {\bf v}\nabla\widehat g
+e{\bf v}{\cal A}\bigl[\widehat\sigma_z \widehat g - 
\widehat g \widehat\sigma_z\bigr]
-{i e \over 2 m}{\cal A}
\bigl[\widehat\sigma_z\nabla\widehat g + \nabla\widehat g 
\widehat\sigma_z\bigr]
\\
\nonumber
&&
+{ie\over 2} \bigl({\bf v}\times {\bf H}\bigr)
	{\partial \over \partial {\bf p}_{\|}}
  \bigl(\widehat\sigma_z \widehat g + \widehat g \widehat\sigma_z \bigr)
+e{\bf v A}
\Bigl[\widehat\sigma_z \widehat g (\omega_n-\omega_0, \omega_{n'})
-\widehat g(\omega_n, \omega_{n'}-\omega_0)\widehat\sigma_z \Bigr]
\\
\label{quasi}
&&
-{i e \over 2 m}
{\bf A}
\Bigl[\widehat\sigma_z \nabla\widehat g (\omega_n-\omega_0, \omega_{n'})
-\nabla\widehat g(\omega_n, \omega_{n'}-\omega_0)\widehat\sigma_z \Biggr]
\\
\nonumber
&&-T\sum_{\omega_k}
     \Bigl[\widehat\sigma(s,{\bf R};\omega_n,\omega_k)
\widehat g(s,{\bf R};\omega_k,\omega_{n'})
- \widehat g(s,{\bf R};\omega_n,\omega_k)
\widehat\sigma(s,{\bf R};\omega_k,\omega_{n'})\Bigr]
\\
\nonumber
&&
+ {i \over 2}\biggl[ {\partial\widehat\Delta\over\partial {\bf R}}
{\partial \widehat g \over\partial{\bf p}_{\|}}
+{\partial \widehat g \over\partial {\bf p}_{\|}}
{\partial\widehat\Delta\over\partial {\bf R}}\biggr]
-{i \over 2}\biggl[ {\partial\widehat\Delta\over\partial {\bf p}_{\|}}
	{\partial \widehat g \over\partial {\bf R}}
+{\partial \widehat g \over\partial {\bf R}}
 {\partial\widehat\Delta\over\partial {\bf p}_{\|}}
\biggl]
\\
\nonumber
&&
-{i \over 2}T\sum_{\bar\omega}
     \biggl[{\partial\widehat\sigma\over\partial {\bf R}}
{\partial \widehat g \over\partial{\bf p}_{\|}}
+{\partial \widehat g \over\partial {\bf p}_{\|}}
{\partial\widehat\sigma\over\partial {\bf R}}\biggr]
-{i \over 2}\biggl[ {\partial\widehat\sigma\over\partial {\bf p}_{\|}}
	{\partial \widehat g \over\partial {\bf R}}
+{\partial \widehat g \over\partial {\bf R}}
 {\partial\widehat\sigma\over\partial {\bf p}_{\|}}
\biggl]
=0,
\end{eqnarray}
where the quasiclassical matrix propagator is, as usual \cite{E,LO}
\begin{equation}
\label{hatgdef}
\widehat g(s,{\bf R}; \omega_n,\omega_{n^{\prime}})=
    \int { d \zeta_p \over \pi}
                \widehat 
                G(p,{\bf R}; \omega_n,\omega_{n^{\prime}})
= 
              \pmatrix{g& -f\cr
                                   f^{\dagger} & \bar{g}\cr}
\end{equation}
and the order parameter is given by the self-consistency condition
\begin{equation}
\label{delta}
\Delta({\bf R})= {\rm g} N(0) \pi \sum_n \int d^2s f(s,{\bf R};
                                                      \omega_n,\omega_n).
\end{equation}
Eqs. (\ref{quasi}) and (\ref{delta}) are 
the generalization of the standard
quasiclassical theory \cite{E,LO} to include terms giving rise to nonzero
Hall conductivity.  

 Before we linearize Eq. (\ref{quasi}) and   solve it
to find the longitudinal and Hall conductivities, several comments should
be made. First, the vector potential of the magnetic field
enters the quasiclassical equation explicitly in contrast to
the case of a normal metal (cf. Eq. (\ref{nbase1})). 
This is readily understood if we notice
that in the last term on the first line
of Eq. (\ref{quasi}) 
the matrix $\widehat\sigma_z\widehat g - \widehat g \widehat\sigma_z$ has only
 off-diagonal
elements, so that the term involving ${\bf v}{\cal A}$ only
appears in equations for the anomalous propagator.
It would seem that the second term involving the vector
potential ${\cal A}$ (the last term in the second line
of Eq. (\ref{quasi})) is present even in a normal metal
since the matrix  
$\widehat\sigma_z\widehat g + \widehat g \widehat\sigma_z$ has only diagonal
elements, and, consequently, this term contributes only to the equations
for the quasiparticle part of the matrix Green's function.
However, in a normal metal in the presence of  uniform
electric and magnetic fields
the response
function is spatially uniform, and this term is irrelevant.
In the superconducting state the spatial variation of the quasiparticle
Green's function is due to the spatial dependence of the
order parameter $\Delta$ in the vortex state, and this term 
describes the coupling of the current, induced by the 
spatial dependence of $\Delta({\bf R})$, to the external field.
Finally, the Lorentz force is accompanied by the matrix propagator
in a combination  
$\widehat\sigma_z\widehat g + \widehat g \widehat\sigma_z$, and therefore the
Lorentz force does not act directly on the Cooper pairs. This result is 
perhaps
not too suprising as in a reference frame associated with the center 
of mass the electrons have opposite momenta, and  
hence there is no net force 
acting on a pair.

\subsection{Linear Response}

We now use the approach given in Section \ref{sec:linmetal}
to linearize the basic equation in the external field.
If we decompose the propagator $\widehat g$, self energy $\widehat\sigma$ and
order parameter $\Delta$ into a leading order term $\widehat g_0$,
$\widehat\sigma_0$ and $\Delta_0$, and a part (denoted by index 1 ) 
linear in the applied electric field,
the equation for the Green's function of a superconductor in a magnetic field
reads
\newpage\noindent
\begin{eqnarray}
\nonumber
&&i \omega_n \bigl(\widehat\sigma_z \widehat g_0 - 
\widehat g_0 \widehat\sigma_z\bigr)
+e{\bf v}{\cal A}\bigl(\widehat\sigma_z \widehat g_0 - 
\widehat g_0 \widehat\sigma_z\bigr)
-\bigl(\widehat\sigma_0
 \widehat g_0
- \widehat g_0
\widehat\sigma_0 \bigr)
+\widehat\Delta_0\widehat g_0 - \widehat g_0 \widehat\Delta_0
\\
\label{g0sup}
&&
+i {\bf v}\nabla\widehat g_0 
+{ie\over 2} \bigl({\bf v}\times {\bf H}\bigr)
	{\partial \over \partial {\bf p}_{\|}}
  \bigl(\widehat\sigma_z \widehat g_0 + \widehat g_0 \widehat\sigma_z \bigr)
-{i e \over 2}{\cal A}
\bigl(\widehat\sigma_z\nabla\widehat g_0 + \nabla\widehat g_0 
\widehat\sigma_z\bigr)
\\
\nonumber
&&
+ {i \over 2}\biggl[ {\partial\widehat\Delta_0\over\partial {\bf R}}
{\partial \widehat g_0 \over\partial{\bf p}_{\|}}
+{\partial \widehat g_0 \over\partial {\bf p}_{\|}}
{\partial\widehat\Delta_0\over\partial {\bf R}}\biggr]
-{i \over 2}\biggl[ {\partial\widehat\Delta_0\over\partial {\bf p}_{\|}}
	{\partial \widehat g_0 \over\partial {\bf R}}
+{\partial \widehat g_0 \over\partial {\bf R}}
 {\partial\widehat\Delta_0\over\partial {\bf p}_{\|}}
\biggl]
\\
\nonumber
&&
-{i \over 2}
     \biggl[{\partial\widehat\sigma_0\over\partial {\bf R}}
{\partial \widehat g_0 \over\partial{\bf p}_{\|}}
+{\partial \widehat g_0 \over\partial {\bf p}_{\|}}
{\partial\widehat\sigma_0\over\partial {\bf R}}\biggr]
+{i \over 2}\biggl[ {\partial\widehat\sigma_0\over\partial {\bf p}_{\|}}
	{\partial \widehat g_0 \over\partial {\bf R}}
+{\partial \widehat g_0 \over\partial {\bf R}}
 {\partial\widehat\sigma_0\over\partial {\bf p}_{\|}}
\biggl]=0,
\end{eqnarray}
while the equation for the response function $g^{(1)}$ is given by
\begin{eqnarray}
\nonumber
&&
i {\bf v}\nabla\widehat g^{(1)}
+ i\omega_n \widehat\sigma_z \widehat g^{(1)} - 
i(\omega_n-\omega_0)\widehat g^{(1)} \widehat\sigma_z
+\widehat\Delta^{(1)}\widehat g_0(-) - \widehat g_0 \widehat\Delta^{(1)}
+\widehat\Delta\widehat g^{(1)}- \widehat g^{(1)}\widehat\Delta
\\
\nonumber
&&
-\bigl[\widehat\sigma_0
 \widehat g^{(1)}
- \widehat g^{(1)}
\widehat\sigma_0(-) \bigr]
-\bigl[\widehat\sigma^{(1)}\widehat g_0(-) 
- \widehat g_0\widehat\sigma^{(1)}\bigr]
+e{\bf v}{\cal A}\bigl(\widehat\sigma_z \widehat g^{(1)}- 
\widehat g^{(1)} \widehat\sigma_z\bigr)
\\
\nonumber
&&
+e{\bf v A}\bigl(\widehat\sigma_z \widehat g_0(-) -\widehat g_0 
\widehat\sigma_z\bigr)
+{ie\over 2} \bigl({\bf v}\times {\bf H}\bigr)
	{\partial \over \partial {\bf p}_{\|}}
  \bigl(\widehat\sigma_z \widehat g^{(1)} + \widehat g^{(1)} 
\widehat\sigma_z \bigr)
\\
\label{g1sup}
&&
-{i e \over 2 m}{\cal A}
\bigl(\widehat\sigma_z\nabla\widehat g^{(1)} + \nabla\widehat g^{(1)} 
\widehat\sigma_z\bigr)
-{i e \over 2 m}
{\bf A}
\bigl(\widehat\sigma_z \nabla\widehat g_0 (-)
-\nabla\widehat g_0\widehat\sigma_z \bigr)
\\
\nonumber
&&
+ {i \over 2}\biggl[ {\partial\widehat\Delta_0\over\partial {\bf R}}
{\partial \widehat g^{(1)} \over\partial{\bf p}_{\|}}
+{\partial \widehat g^{(1)} \over\partial {\bf p}_{\|}}
{\partial\widehat\Delta_0\over\partial {\bf R}}\biggr]
+ {i \over 2}\biggl[ {\partial\widehat\Delta^{(1)}\over\partial {\bf R}}
{\partial \widehat g_0(-) \over\partial{\bf p}_{\|}}
+{\partial \widehat g_0 \over\partial {\bf p}_{\|}}
{\partial\widehat\Delta^{(1)}\over\partial {\bf R}}\biggr]
\\
\nonumber
&&
-{i \over 2}\biggl[ {\partial\widehat\Delta_0\over\partial {\bf p}_{\|}}
	{\partial \widehat g^{(1)} \over\partial {\bf R}}
+{\partial \widehat g^{(1)} \over\partial {\bf R}}
 {\partial\widehat\Delta_0\over\partial {\bf p}_{\|}}
\biggl]
-{i \over 2}\biggl[ {\partial\widehat\Delta^{(1)}\over\partial {\bf p}_{\|}}
	{\partial \widehat g_0(-) \over\partial {\bf R}}
+{\partial \widehat g \over\partial {\bf R}}
 {\partial\widehat\Delta^{(1)}\over\partial {\bf p}_{\|}}
\biggl]
\\
\nonumber
&&
-{i \over 2}
     \biggl[{\partial\widehat\sigma_0\over\partial {\bf R}}
{\partial \widehat g^{(1)} \over\partial{\bf p}_{\|}}
+{\partial \widehat g^{(1)} \over\partial {\bf p}_{\|}}
{\partial\widehat\sigma_0(-)\over\partial {\bf R}}\biggr]
-{i \over 2}
     \biggl[{\partial\widehat\sigma^{(1)}\over\partial {\bf R}}
{\partial \widehat g_0(-) \over\partial{\bf p}_{\|}}
+{\partial \widehat g_0 \over\partial {\bf p}_{\|}}
{\partial\widehat\sigma^{(1)}\over\partial {\bf R}}\biggr]
\\
\nonumber
&&
+{i \over 2}\biggl[ {\partial\widehat\sigma_0\over\partial {\bf p}_{\|}}
	{\partial \widehat g^{(1)} \over\partial {\bf R}}
+{\partial \widehat g^{(1)} \over\partial {\bf R}}
 {\partial\widehat\sigma_0(-)\over\partial {\bf p}_{\|}}
\biggl]
+{i \over 2}\biggl[ {\partial\widehat\sigma^{(1)} \over\partial {\bf p}_{\|}}
	{\partial \widehat g_0(-) \over\partial {\bf R}}
+{\partial \widehat g_0 \over\partial {\bf R}}
 {\partial\widehat\sigma^{(1)} \over\partial {\bf p}_{\|}}
\biggl]=0,
\end{eqnarray}
The rest of this work will
be devoted to solving these two equations to determine the transverse 
electrical
 conductivity of a type-II superconductor in the vortex state.

To calculate the response of
a superconductor it will be
 convenient to modify the definition of current given in
Eq. (\ref{current}). Since the diagonal elements of the matrix
propagator are related by Eq. (\ref{g22}), it is easy to check that the
current  can be written as
\begin{equation}
\label{supcurr}
{\bf j}({\bf R})= {1\over 2}\pi e N(0)
\sum_\omega \int d^2 s   {\bf v} (s)  (g_1-\bar g_1),
\end{equation}
where $g_1$ and $\bar g_1$ are the diagonal elements of the
response function $\widehat g^{(1)}$. 
In the standard quasiclassical approach
the distribution function $g$ also satisfies a 
``normalization condition'' \cite{E,LO}
\begin{equation}
\label{norma}
\sum_{\omega_k} \widehat g(\omega_n,\omega_k)
\widehat g(\omega_k, \omega_{n'})=-\delta_{\omega_n,\omega_{n'}}.
\end{equation}
In particular, using this condition for the leading order distribution
function, which is diagonal in frequency (see Eq. (\ref{g1sup}))
we find
\begin{equation}
\label{normag0}
\widehat g_0^2(\omega_n)=-1.
\end{equation}
However, it has to be emphasized that this normalization condition 
holds if
and only if the gradient of the function $g$ can be written as
a commutator of an operator with the distribution function,
as is evident from the original derivation \cite{E,LO}.
It does not 
apply when terms responsible for the Hall effect are taken into account,
since they have the form of an anticommutator of a matrix operator
with the Green's function. Nevertheless,
this normalization condition will prove useful in determining the 
quasiclassical Green's function of a superconductor in 
a high magnetic field at zeroth order. 
 
\section{Type-II Superconductor in a High Magnetic Field}
\label{chap:dos}

\subsection{Model}

We consider a clean type-II superconductor in a magnetic field $H$
close to the upper critical field $H_{c2}$.
Again we consider a spherical Fermi surface, and 
impurity scattering is treated in the Born approximation.
 The condition for a superconductor
to be in the clean regime 
is $l \gg \xi_0$. 
In fields not too far below the upper critical field
the magnetic length $\Lambda \simeq \xi_0$, so that in the clean
regime $l \gg \Lambda$.
In type-II superconductors the spatial variations
of the internal field become less pronounced as the superfluid 
density decreases with increased applied uniform magnetic field. As a result,
 near $H_{c2}$
internal fields can be
assumed spatially uniform and equal to the applied field and
the vortex lattice 
can be modeled by an order parameter of the same form as the
periodic Abrikosov solution \cite{abrik}
\begin{equation}
\label{VL}
\Delta({\bf R})=\sum_{k_y} C_{k_y} e^{i k_y y} 
\exp\bigl(-(x - \Lambda^2 k_y)^2
                          / 2 \Lambda^2 \bigr)
=\sum_{k_y} C_{k_y} e^{i k_y y} \Phi_0 (x - \Lambda^2 k_y),
\end{equation}
where  $\Phi_0 (x)$ is the lowest energy eigenfunction of the linearized
Ginzburg-Landau Eq. (i.e. the eigenfunction of a harmonic oscillator
with the Cooper pair mass $M=2m$ and frequency $\omega_c$). The vector
potential of the magnetic field has been chosen in an asymmetric gauge
${\cal A}({\bf R})= ( 0, Hx, 0)$.
The periodicity of the coefficients $C_{k_y}$ determines the type of 
the vortex lattice. 
Here, we do not consider a specific periodicity, the only
assumption made is that there are flux lines in the system; this 
solution, therefore, can serve as a model for a rigid line
liquid as well.

\subsection{Quasiclassical Equations in the Absence of an Electric Field
and the BPT Approximation}

First we consider the leading order Eqs. (\ref{g0sup}) and
neglect  terms of order $\omega_c$.
Then the elements are
\begin{eqnarray}
\label{eqg0}
   i {\bf v} \nabla g + \Delta f^{\dagger} - \Delta^{\star} f
         &=&{1 \over 2 \tau} \Delta \langle f^{\dagger} \rangle -
          {1 \over 2 \tau} \Delta^{\star}\langle f \rangle
\\
\label{eqf0}
(2 \omega_n  + {\bf v}( \nabla - 2 i e {\cal A})) f&=&
2 i \Delta g + {i \over \tau} \langle f \rangle g
-{i \over \tau} \langle g \rangle f
\\
\label{eqfd}
(2 \omega_n  - {\bf v}( \nabla + 2 i e {\cal A})) f^{\dagger}&=&
2 i \Delta^{\star}  g + {i \over \tau} \langle f^{\dagger}
\rangle g
-{i \over \tau} \langle g \rangle f^{\dagger},
\end{eqnarray}
here angular brackets denote an average over the Fermi surface.
The normalization condition, Eq. (\ref{norma}), 
can be used in this case, so that
\begin{eqnarray}
\label{ng0}
g^2 - f f^{\dagger}& =& - 1
\\
\label{ngbar}
g + \bar {g} &=&0.
\end{eqnarray}

To solve these equations we employ the approach due
to Brandt, Pesch, and Tewordt \cite{BPT}, which was first used in the
framework of the quasiclassical approximation by Pesch \cite{Pesch,KP}.
In this method the diagonal elements $g$ and
$\bar g$ of the
matrix propagator are approximated 
by their spatial averages, while
the exact spatial form of $\Delta({\bf R})$ is retained 
in determining the off diagonal
functions $f$ and $f^{\dagger}$. The crucial observation 
is that  the diagonal
part of the Green's function is periodic in the 
center of mass coordinate ${\bf R}$ with the
same periodicity as the order parameter.
Performing a Fourier decomposition of the full Green's function in 
the vectors
${\bf K}$ of the reciprocal flux line lattice, these authors
 \cite{BPT} showed that the Fourier components 
of the Green's function with
${\bf K} \neq 0$ are exponentially small (by a factor
 $\exp( - \Lambda^2 K^2)$)
compared to the component with  ${\bf K}=0$.
This component is, of course, the spatial average
of the Green's function over a unit cell of the vortex lattice,
which suggests the above approximation. 

The diagonal part of the distribution function
depends on the amplitude of the order parameter, but not on its phase.
 The length scale for the suppression of the mean 
field order parameter
amplitude by
a single vortex is the coherence length $\xi_0$, therefore 
near the
upper critical field
 the 
order parameter is globally suppressed in the bulk
of the superconductor. Consequently, spatial variations
of the amplitude $|\Delta|^2$ can be ignored for fields
close to $H_{c2}$.
On the other hand, as the phase of the order parameter changes
by $2\pi$ around a single vortex, the rapid spatial
variation of phase in the vortex state must be
taken into account to determine
 the
off diagonal elements of the quasiclassical propagator.
After averaging over a single unit cell, the remaining
spatial dependence of the amplitude   $|\Delta|^2$
is determined by the nonuniformity of the electromagnetic fields;
the relevant length scale is the London penetration depth $\lambda_L$.
 Therefore, the BPT
approximation 
works very well for superconductors in the London limit
$\kappa=\lambda_L / \xi_0 \gg 1$; even 
for materials with moderate values of $\kappa$
 it remains valid over a wide field range below $H_{c2}$.
Numerical results obtained by Brandt \cite{brandthesis} indicate
that the BPT approximation works extremely well as long
as the parameter $(\Lambda \Delta /v) \leq 0.3$. Since the 
field dependence of the magnetic length is slow
$\Lambda\approx \xi_0 (H_{c2}/H)^{1/2}$, this means that
the approximation can be used over almost  the entire
region of linear magnetization, where the order parameter is suppressed.

In all of the following $g$ stands for
the spatially averaged distribution function. 
To  determine the functions $g$, $f$ 
and $f^{\dagger}$ we solve Eqs. (\ref{eqf0}) and (\ref{eqfd}) for
the off diagonal elements of the matrix distribution function in
terms of $g$, and apply the spatially averaged normalization condition
of Eq. (\ref{ng0}) to determine the diagonal part self-consistently.
We introduce the impurity renormalized frequency
\begin{equation}
\label{wtilde}
\widetilde\omega_n= \omega_n + {i \over 2\tau}\langle g(\widetilde\omega_n)
\rangle
\end{equation}
and rewrite the equations for the off diagonal part of the distribution
function as
\begin{eqnarray}
\label{eqf}
&&f=\bigl(2\widetilde\omega_n + {\bf v}( \nabla - 2 i e {\cal A})\bigr)^{-1}
\bigl(2ig\Delta +{i \over \tau} \langle f \rangle g \bigr)
\\
\label{eqfd1}
&&f^{\dagger}=\bigl(2\widetilde\omega_n -{\bf v}( \nabla +
2 i e {\cal A})\bigr)^{-1}
\bigl(2ig\Delta^{\star}+{i \over \tau} \langle f^{\dagger}\rangle g \bigr).
\end{eqnarray}
To proceed with  this program we need to know the
result of acting with the operator 
$\bigl(2\widetilde\omega_n \pm {\bf v}( \nabla \mp 2 i e {\cal A})\bigr)^{-1}$
on the order parameter. 

\subsection{Operator Formalism}

Since the order parameter given in
Eq. (\ref{VL}) is a superposition of the lowest energy 
eigenfuctions of a harmonic oscillator centered at 
different vortex cores, we  introduce the
raising and lowering operators
\begin{eqnarray}
\label{opa}
a&=&
         {\Lambda \over \sqrt2}[\nabla_x + i ( \nabla_y -2 i e H x)]
\\
\label{opad}
a^{\dagger}& =& 
          -{\Lambda \over \sqrt2}[\nabla_x - i ( \nabla_y -2 i e H x)].
\end{eqnarray}
These operators obey the 
usual bosonic commutation relations
 $[a,a^{\dagger}]=1$.
We now interpret the Abrikosov solution as the ground state of this
ensemble  of oscillators $\Delta=|0\rangle$. The higher eigenstates
of the system are generated by the standard formula
\begin{equation}
\label{excited}
 a^{\dagger}|n\rangle =\sqrt{n+1}|n+1\rangle,
\end{equation}
this operation excites oscillator states centered on each vortex line, so that
\begin{equation}
\label{excite}
|n\rangle =\sum_{k_y} C_{k_y} e^{i k_y y} \Phi_n (x - \Lambda^2 k_y)
\end{equation}
Similarly we can  introduce conjugate operators corresponding to
$\Delta^{\star}=\langle 0|$, the raising and lowering operators
for these states 
are now defined as  $b=(a)^{\star}$ and $b^{\dagger}=(a^{\dagger})^{\star}$.
Wide use of bosonic operators for the description of the vortex lattice 
has been 
hampered by the fact that, even though
the wavefunctions corresponding to different oscillator
states centered on the same vortex line are orthogonal,
functions centered on different flux lines
overlap, so that different excited states as defined above are not orthogonal
and the equations are non-local
(see, for example, Ref. \cite{E1}). What makes this approach successful 
when combined with the BPT approximation is that
this set of states is orthogonal in the sense of a spatial
average
\begin{equation}
\label{ortho}
\int d^3 R \langle m|n \rangle = \Delta^2 \delta_{m,n},
\end{equation}
where $\Delta$ is the spatial average of the order parameter.
This condition is obeyed since the phase factor $\exp(i k_y y)$
ensures that only functions centered on the same site contribute to the
integral. 
Therefore if we are only concerned  with spatial averages of physical
quantities, the excited states of the order parameter can be treated 
as states of a harmonic oscillator.

To evaluate the result of acting with the gradient operator 
${\bf v}(\nabla - 2ie{\cal A})$ on the order parameter $\Delta$
we rewrite it
in terms of  the raising and lowering operators $a$ and $a^{\dagger}$
\begin{equation}
{\bf v}(\nabla - 2ie{\cal A})=
{v \sin \theta \over \sqrt2 \Lambda} 
[ a e^{-i \phi} - a^{\dagger} e^{i \phi}].
\end{equation}
Then the result of the action 
of the operator 
$\bigl(2\widetilde\omega_n + {\bf v}( \nabla - 2 i e {\cal A})\bigr)^{-1}$
on any mode $|m\rangle$ of the order parameter 
can be evaluated  exactly. The technical details
are given in Appendix \ref{sec:operators}, here we give only the
final result
\begin{equation}
\label{opmain}
(2 \widetilde \omega_n +
{\bf v}  ( \nabla - 2 i e {\cal A}) )^{-1} | m \rangle  =
{\sqrt{\pi} \Lambda \over v \sin \theta}
            \sum_{m_2=0}^{\infty} \sum_{m_1=0}^m 
              D_m^{m_1m_2} e^{i(m_2 -m_1) \phi} 
              | m+m_2-m_1 \rangle,
\end{equation}
 where 
\begin{eqnarray}
\label{opd}
&&D_m^{m_1m_2}=
        {\sqrt{m!} \sqrt{(m-m_1+m_2)!} \over (m-m_1)! m_1! m_2!}
          (-1)^{m_1}(-{i \over \sqrt2})^{m_1+m_2}
\bigl({\rm sgn}(\omega_n) \bigr)^{m_1+m_2+1}                  
                        W^{(m_1+m_2)} (u_n),
\\
\label{un}
&& u_n= {2 i \widetilde \omega_n \Lambda{\rm sgn}
                                     (\omega_n)
                                     \over
                                       v \sin \theta},
\\
\label{W}
&&
W(u)=e^{-u^2}{\rm erfc}(-iu),
\end{eqnarray}
and $W^{(m)}$ is the $m$-th derivative of the function $W$.
Eq. (\ref{opd}) is the main result of the operator formalism 
developed here; it allows further
progress towards a solution of the 
quasiclassical equations to be made.

\subsection{Type-II Superconductor in High Magnetic Field}
\label{sec:ssolution}

Guided by the work of Eilenberger \cite{E1} and Pesch \cite{Pesch}, 
we make an ansatz solving Eq. (\ref{eqf}) for an s-wave
superconductor. This ansatz makes use of the fact that
the term dependent on impurity scattering 
in the right hand side of the equation 
renormalizes the amplitude of the order parameter
\begin{equation}
f=2i g D^{-1}(\widetilde\omega_n)
\Bigl(2\widetilde\omega_n + {\bf v}\bigl(\nabla-2ie{\cal A}\bigr)\Bigr)^{-1}
\Delta.
\label{ans}
\end{equation}
Since the order parameter $\Delta$ in this equation is the
``ground state'' of the Abrikosov vortex lattice $|0\rangle$, the form of the
function $f$ can be  obtained immediately from 
Eqs. (\ref{opmain}) and (\ref{oper0})
\begin{equation}
\label{f}
\qquad
f(s)=2i g(s) D^{-1}(\widetilde\omega_n)
{\sqrt{\pi} \Lambda \over v \sin \theta}
 \sum_{m=0}^{\infty} {1 \over \sqrt{m!}} (-{i \over \sqrt2})^m e^{i m \phi}
              ({\rm sgn}(\omega_n))^{m+1}
             W^{(m)} (u_n) | m \rangle.
\end{equation}
Substituting this expression into Eq. (\ref{eqf}), we 
find for the impurity renormalization of the order parameter
\begin{equation}
D(\widetilde\omega_n)=
1-i\sqrt\pi {\Lambda\over 2l}{\rm sgn}(\omega_n)
\int_0^\pi d\theta g(\theta;\widetilde\omega_n)W(u_n).
\label{denom}
\end{equation}
Using the corresponding Eq. (\ref{eqfd1}) we obtain $f^{\dagger}(s)$
\begin{equation}
\label{fdagger}
\qquad
f^{\dagger}(s)=2i g(s) D^{-1}(\widetilde\omega_n)
{\sqrt{\pi} \Lambda \over v \sin \theta}
 \sum_{m=0}^{\infty} {1 \over \sqrt{m!}} ({i \over \sqrt2})^m e^{-i m \phi}
              ({\rm sgn}(\omega_n))^{m+1}
             W^{(m)} (u_n)\langle m |.
\end{equation}

Then we can use the normalization condition,
Eq. (\ref{ng0}), to determine $g$ 
(see Appendix \ref{sec:operators} for
details)
\begin{equation}
\label{gs}
g=- {i }{\rm sgn}(\omega_n ) P(\theta,\widetilde\omega_n),
\end{equation}
where
\begin{equation}
\label{P}
 P(\theta,\widetilde\omega_n)=
                   \Bigl[ 1- i \sqrt{\pi} 
                      \bigl({2 \Lambda \Delta \over D v \sin\theta}\bigr)^2
                           W^{\prime}
                    \bigl (u_n)
                                 \Bigr]^{-1/2},
\end{equation}
and the sign has been chosen to give the correct expression 
in the normal
state.
Eqs. (\ref{wtilde}) and  (\ref{f}) - (\ref{P}) provide a complete
self consistent solution of the quasiclassical equations 
for an s-wave superconductor in a magnetic field. 
A Green's function very similar to that given in Eqs. (\ref{gs})
and (\ref{P})
was obtained in the work of Pesch \cite{Pesch}
by a different method. 
As in the microscopic theory, the order parameter is determined from the
self-consistency condition given by Eq. (\ref{delta}),
which is in this case
\begin{equation}
\label{sgapsph}
1=i \pi \sqrt\pi {\rm g} N(0) {\Lambda\over v}
\sum_n \int_0^\pi d\theta  g(\widetilde\omega_n)D^{-1}(\widetilde\omega_n)
		{\rm sgn}(\omega_n) W(u_n).
\end{equation}

For the general case of finite mean free path and applied magnetic
field a closed form solution of the self-consistent expressions
cannot be easily found. 
However, with minor simplifications it is possible
to obtain analytical results from this solution.
Even though the dimensionless parameter 
$(\Lambda\Delta /v)^2$
in the 
Green's function given
by Eq. (\ref{gs}) is small
in the region where the BPT approximation is valid, it appears
 with the weight
$(\sin\theta)^{-2}$, so that a straightforward expansion
is impossible. We will see, in fact, that the density of states
is a non-analytic function of this parameter. 
However,
 while the
full functional dependence of the Green's function
on  $(\Lambda\Delta /v)^2$ has to be  retained, terms of higher
order in this small quantity can be neglected in this functional
form, provided that they do not result in more singular behaviour.
Both the impurity renormalization of  the  frequency $\widetilde\omega_n$ and 
the renormalization of the order parameter depend on the weighted
angular average of the Green's function $g$, which is non-singular as a
function of  the
order parameter in the vortex state, 
this is related to the gapless character of the quasiparticle
spectrum. 
Therefore, in determining the function $P$ to leading
order in $(\Lambda\Delta /v)^2$ 
the Green's function in the definition of impurity renormalization 
of the order parameter
$D$ (Eq. (\ref{denom})) 
can be replaced by its normal state value.
Similarly, the renormalized frequency in the argument of the
function $W'$  can be replaced by
$\omega_n+{\rm sgn}(\omega_n)/2\tau$. 
The resulting expression for
the renormalization function is identical to that
obtained by Helfand and Werthamer \cite{helfand}.
 With these approximations, Eqs. (\ref{f})-(\ref{P}) 
describe a closed form solution.
In the clean limit near the upper critical field of interest here
expressions for the quasiclassical propagator can be simplified even 
further. Since in this regime
$l\gg\Lambda$, and the renormalization of the order parameter
is $D=1+O(\Lambda/l)$, to leading order $D\approx 1$.

The anomalous Green's functions
$f$ and $f^{\dagger}$ are given as a Fourier series in the azimuthal angle
$\phi$, with the $m$-th component of the series coupling to the
$m$-th excited state (or mode) of the order parameter $\Delta$.
Therefore in the presence of an external perturbation
the mode with $m=0$ will couple to a scalar potential,
the mode with $m=1$ to a transverse potential etc.
The function $P$ given in Eq. (\ref{P}) is related to the angular dependent
density of states. If the Green's function is analytically continued into the 
upper half plane by letting $i\omega_n\rightarrow\omega+i\delta$
then the density of states 
\begin{equation}
N(\omega,\theta)=
-N(0) {\rm Im} g(\omega,\theta)=
N(0) {\rm Re} P(\omega,\theta)
\end{equation}
is strongly angular dependent.
For quasiparticles travelling parallel
to the magnetic field, $N(\omega,\theta)$ is gapped and BCS like,
 while in all other
directions it is gapless. The total density of states (DOS) $N_s(\omega)$, 
obtained by angular integration
of the imaginary part of the function $g$, is gapless \cite{BPT},
while the residual density of states at the Fermi surface $N_s(0)$
is a non-analytic function of the order parameter \cite{BPT,Pesch}
\begin{equation}
\label{resdos}
N_s\approx N(0)\Bigl[1-4\Bigl({\Lambda\Delta\over v}\Bigr)^2
		\ln\Bigl({\sqrt2 v \over\Lambda\Delta}\Bigr)
		+2\Bigl({\Lambda\Delta\over v}\Bigr)^2\Bigr].
\end{equation}
The Green's function obtained here also
reproduces
the BCS Green's function if the limit $H\rightarrow 0$ is taken,
which suggests that it can be used to interpolate between the high-field and
the low field regimes.
We now have a closed form 
expression
for the matrix propagator near the upper critical field up to the
order $(\Lambda\Delta /v)^4$, which we will use
to determine the 
linear response of a superconductor to an electric field.

\section{LONGITUDINAL CONDUCTIVITY}
\label{chap:sxx}

We begin by considering  the longitudinal conductivity in the vortex
state in the BPT approximation. 
We are concerned here with the transport coefficients
in the clean limit, and will neglect all contributions
to conductivity of relative order $(\Lambda/l)$ compared to
the most significant modifications upon entering the
superconducting state. 
We again omit
terms of order of cyclotron frequency. 

\subsection{The Response Function}

Since the electrical 
current given in Eq. (\ref{supcurr}) depends on 
$g_e=g_1-\bar g_1$, we write the linearized quasiclassical Eq. (\ref{g1sup})
for 
the spatial average of this combination.
Then the equations for the linear, in the applied electric field, 
averaged diagonal elements of the
distribution function, and the equations for the anomalous functions are
\begin{eqnarray}
\label{Gel}
g_e=g_1-\bar g_1&=&{2e{\bf v}{\bf A} (g - g(-))\over i \widetilde\omega_0}
+ (i \widetilde\omega_0)^{-1} \bigl( \overline{\Delta_1^{\star} f}- 
                   \overline{ \Delta_1^{\star}f(-)}\bigr)+ 
   (i \widetilde\omega_0)^{-1}\bigl(   \overline{   \Delta_1 f^{\dagger}}- 
		\overline {\Delta_1 f^{\dagger}(-)}\bigr)
\\
\nonumber
&&
+(2i \widetilde\omega_0\tau)^{-1} \biggl(\bigl( 
 \overline{ \langle f_1^{\dagger} \rangle f}-
 \overline{ \langle f_1^{\dagger} \rangle f(-)}\bigr) + 
 \bigl(\overline{\langle f_1 \rangle f^{\dagger}}- 
\overline{\langle f_1 \rangle f^{\dagger}(-)}\bigr)
					\biggr)
\\
\nonumber 
&&
-(2i \widetilde\omega_0\tau)^{-1} \biggl(\bigl(
 \overline { f_1^{\dagger}\langle f\rangle }-
 \overline { f_1^{\dagger}\langle f(-)\rangle }\bigr)
		+
\bigl( \overline { f_1\langle f^{\dagger}\rangle}- 
\overline { f_1\langle f^{\dagger}(-)\rangle}\bigr)
						\biggr)
\end{eqnarray}
\begin{eqnarray}
\label{f1s}
\Bigl[ 2 \widetilde \Omega_n + {\bf v} (\nabla - 2 i e {\cal A})\Bigr] f_1 
         & =&
            i e {\bf v} {\bf {A}} (f + f(-))
	+ i \Delta (g_1 - \bar g_1) +
                          i \Delta_1 (g + g(-))
\\
\nonumber
&&
 		+ i (2\tau)^{-1}\Bigl(
		\langle f_1 \rangle (g+g(-))-\langle f \rangle \bar g_1
			+\langle f(-)\rangle g_1 \Bigr)
\\
\label{f1ds}
\Bigl[ 2 \widetilde \Omega_n - {\bf v} (\nabla +2 i e {\cal A})\Bigr] 
					f_1^{\dagger} 
         & =&
            i e {\bf v} {\bf {A}} (f^{\dagger} + f^{\dagger}(-))
+ i \Delta^{\star} (g_1 - \bar g_1) +
                          i \Delta_1^{\star} (g + g(-))
\\
\nonumber
&&
+ i (2\tau)^{-1}
\Bigl(\langle f_1^{\dagger} \rangle (g+g(-))-
\langle f^{\dagger}(-) \rangle \bar g_1
			+\langle f^{\dagger}\rangle g_1 \Bigr).
\end{eqnarray}
The notation used here is identical to that of
the previous Section, and the frequency $\widetilde\Omega_n$ is defined as
%\begin{equation}
$\widetilde\Omega_n=\widetilde\omega_n + \widetilde\omega-$.
%\end{equation}
It is possible to identify the different contributions to
the right hand side of Eq. (\ref{Gel}). The first term 
is the quasiparticle contribution to the current, this term 
determines the response function in the normal state, and, 
with the modified Green's function,
describes the contribution of quasiparticles to the current in superconductors.
The other terms on the right hand side exist only in the superconducting
state. The first two of these involve the modification of the order
parameter $\Delta$, and can be associated with the motion of the 
vortex lattice
under the influence of the applied electric field.   
The remaining terms mix the contributions of the quasiparticles and
the Cooper pairs. It will be shown below that the most relevant
contribution from these terms is due to the additional scattering of
the quasiparticles by dynamical fluctuations of the order parameter, 
similar to the processes described  in the dirty limit by 
the Thompson diagrams
\cite{thompson}.

The
quasiparticle contribution to
the response function $g_1-\bar g_1$ 
can be determined immediately since the
unperturbed functions $g$ and $g(-)$ are known from Eq. (\ref{gs}). 
To evaluate the other contributions to the response function,
Eqs. (\ref{f1s}) and (\ref{f1ds}) have to be solved for
$\Delta_1$ and  $\langle f_1 \rangle$, as well as for the 
conjugate quantities
$\Delta_1^{\star}$ and $\langle f_1^{\dagger} \rangle$.
As before, here we determine 
the complete functional dependence 
of the response function on the order parameter to
order $\Delta^2$, and neglect corrections that
vanish faster than this  as $\Delta$ decreases.
Since both $\Delta_1$ and $f^{\dagger}$ can be 
expanded in
a complete set of functions $|m\rangle$ and $\langle m|$, 
which are normalized by $\Delta^2$, see Eq. (\ref{ortho}),
 it is sufficient 
to determine the  expansion coefficients  to 
zeroth
order in $\Delta$. Therefore in Eqs. (\ref{f1s}) and (\ref{f1ds})
we can replace the
functions $g$,  $g_1$ and $\bar g_1$ by their normal state
values. 
With these simplifications Eqs. (\ref{Gel})-(\ref{f1ds})
can be solved explicitly for
$\Delta_1$, $f_1$ and the  ``daggered'' functions.

\subsection{Quasiparticle Contribution}
\label{sec:sxxqp}

Two different effects modify the quasiparticle contribution to the
current
relative  to the current in a normal metal. First, the 
 difference $g-g(-)$ is
modified relative to  its normal state form, and,
second, as
the impurity renormalization of the frequency $\widetilde\omega_0$ depends
on the unperturbed Green's function it is also affected by the opening
of the superconducting gap below the upper critical field.
In the normal state,
the difference $g-g(-)$ vanishes in the outside frequency region,
for a type-II superconductor
this difference is of order $\Delta^2$.
Therefore in a calculation to lowest order in $\Delta^2$  
the renormalized frequency 
can be replaced by the bare frequency in this frequency range. 
 On the other hand, in the 
intermediate frequency range, where the difference of the unperturbed Green's
functions $g-g(-)$ is of order 1, it is important to keep the full dependence
of the renormalized frequency on the order parameter.
Further, as the contribution from the outside
region is proportional to $\bar\omega$, but not $\tau$,
it is of order $(\Lambda\Delta/v)^2(\Lambda/l)$ and negligible
compared to the contribution from the intermediate frequency range.
 This situation is not unusual when comparing different
  contributions to
the conductivity. Two dimensionless quantities
involving the frequency of the external 
electric field appear in our analysis.
The first, $\bar\omega\tau$,  usually comes from renormalization
of the bosonic frequency $\omega_0$ in the intermediate frequency range.
The second, $(\Lambda\bar\omega/v)$,  appears 
 when the response functions are expanded in the external frequency 
since the argument $u_n$ of these
functions involves the  frequency in the combination 
$(\Lambda\bar\omega/v)$, see Eq. (\ref{un}).
In the dc response only  terms linear in $\bar\omega$ contribute to the
absorptive part of the conductivity. Therefore, as the ratio of the two
dimensionless parameters is of order $(\Lambda/l)$, we  keep 
terms of order $\bar\omega\tau$ while neglecting 
those of order $(\Lambda\bar\omega/v)$.
In the quasiparticle contribution then the only relevant
terms arise from the intermediate frequency range.

Since the quasiparticle spectrum is gapless in the high field
regime, the response function varies slowly over the scale $\omega\sim T$,
and the frequency sums can be evaluated easily, see Appendix \ref{sec:sums}.
We find
 the quasiparticle contribution to the  current
\begin{equation}
\label{jqpfull1}
{\bf j}_{qp}={1\over 4}N(0)e^2 v^2 {\bf A} \int_0^\pi \sin^3\theta d\theta
\Biggl[(P-1) +i\bar\omega\tau\Bigl[P + \langle (1-P)\rangle
-\sqrt\pi W''\Bigl({\Lambda\over l\sin\theta}\Bigr)
\Bigl({2\Lambda\Delta\over v\sin\theta}\Bigr)^2
\Bigl({\Lambda\over l\sin\theta}\Bigr)P^3\Bigr]\Biggr],
\end{equation}
here all the functions are evaluated at $\omega=0$. For $\omega=0$ the argument
of the function $W'$ in Eq. (\ref{P}) is  purely imaginary, 
and the function
$P$ is purely real.
It follows that the  first term in Eq. (\ref{jqpfull1}) 
contributes to the  non-absorptive
part of the conductivity; it is the remnant of the  Meissner effect 
in a type-II
superconductor in a magnetic field.
The remaining terms contribute to the absorptive part, and
the transport current can be written as 
\begin{equation}
\label{jqpfull}
{\bf j}_{qp}={1\over 4}N(0)e^2 v^2\tau  {\bf E}
\int_0^\pi \sin^3\theta d\theta
\Biggl[\Bigl[P-1\Bigr] + \Bigl[1+\langle (1-P)\rangle\Bigr]
-\sqrt\pi W''\Bigl({\Lambda\over l\sin\theta}\Bigr)
\Bigl({2\Lambda\Delta\over v\sin\theta}\Bigr)^2
\Bigl({\Lambda\over l\sin\theta}\Bigr)P^3\Biggr].
\end{equation}
 The first term in Eq. (\ref{jqpfull}) is
the direct modification of the quasiparticle current on entering
the superconducting state
\begin{equation}
\label{jqpdos1} 
{\bf j}_{qp1}={1\over 4}N(0)e^2 v^2 \tau {\bf E} 
\int_0^\pi \sin^3\theta d\theta\Biggl[
\Bigl[ 1- i \sqrt{\pi} 
                          \bigl({2 \Lambda \Delta \over v \sin\theta}\bigr)^2
                           W^{\prime}
                     (i\Lambda/l\sin\theta)
                                 \Bigr]^{-1/2}-1\Biggr]
={\bf j'}_{qp} - \sigma_n {\bf E},
\end{equation}
where the normal state conductivity $\sigma_n$ was defined in Section 
\ref{sec:normalsigma}. The contribution of small angles
$\sin\theta\leq(\Lambda/l)$ to the angular integrals is
of higher order in $(\Lambda/l)$ 
and can be neglected. For larger angles the argument of the function
$W^{\prime}$ can be set to zero since $\Lambda/l\ll 1$. 
Then the integration is easily carried out,
expanding the resulting elliptic integrals for small values of 
the parameter $(\Lambda\Delta/v)$, we find that
 the correction to the 
conductivity from this term 
\begin{equation}
\label{sxxqp1}
\Delta\sigma_{xx}^{qp1}=-6\sigma_n\Bigl({\Lambda\Delta\over v}\Bigr)^2
\end{equation}
is negative. In the superconducting state in addition
to the  scattering of  quasiparticles by impurities, 
quasiparticles are scattered
by the vortex lattice. At a vortex core
a quasiparticle can undergo
Andreev scattering into a hole and a Cooper pair with no
energy cost. This additional scattering process reduces the
quasiparticle contribution to the current. 

The second term in Eq. (\ref{jqpfull}) arises from 
renormalization of the scattering time $\tau$ in the vortex state.
It can be written as
\begin{equation}
\label{jqptau}
{\bf j}_{qp2}=\sigma_n\biggl[1+\Bigl(1-\langle P\rangle\Bigr)\biggr]{\bf E}
={1\over 3} N(0) e^2 v^2\tau_{eff}{\bf E},
\end{equation}
where the scattering rate 
\begin{equation}
\label{taueff}
\tau_{eff}=\tau \biggl[1+\Bigl(1-\langle P\rangle\Bigr)\biggr].
\end{equation}
The quantity $N(0)\langle P\rangle$ evaluated
at $\omega=0$ is the residual density of states in a superconductor $N_s$,
see Eq. (\ref{resdos}).
Hence this
term describes the effect of the change in the density of states on
the scattering rate of the quasiparticles. 
Below the transition,
 as the superconducting gap opens,
the residual density
of states at the Fermi surface is suppressed compared to 
the density of states
in the normal state;
consequently, the effective scattering rate is smaller and
the effective mean free path is larger. The angular integral of
the function $P$ can be evaluated to leading order in
$(\Lambda/l)$ and expanded in $(\Lambda\Delta/v)$
in similar fashion to the integral analyzed above, we obtain,
in agreement with the result of Eq. (\ref{resdos}),
the
effective scattering time 
\begin{equation}
\label{taueffective}
\tau_{eff}=\tau\biggl[1 - 4 \biggl({\Lambda \Delta \over v}\biggr)^2
     	\ln \biggl({\Lambda \Delta \over \sqrt 2 v}\biggr)
	- 2 \biggl({\Lambda \Delta \over v}\biggr)^2 \biggr],
\end{equation}
and  the contribution to the longitudinal conductivity 
\begin{equation}
\label{sxxqptau}
\sigma_{xx}^{qp2}=\sigma_n \Bigl[1+4\Bigl({\Lambda\Delta\over v}\Bigr)^2
		\ln\Bigl({\sqrt2 v \over\Lambda\Delta}\Bigr)
		-2\Bigl({\Lambda\Delta\over v}\Bigr)^2\Bigr].
\end{equation}
Since $(\Lambda\Delta/v)\ll 1$, the logarithmic term dominates 
near the transition and this contribution is enhanced
relative to the normal state value. The last term in Eq. (\ref{jqpfull})
contributes at order $(\Lambda\Delta/v)^2(\Lambda/l)$.

\subsection{Dynamical Fluctuations of the Order Parameter}
\label{sec:sxxfl}

To compute the contribution of all the other terms
in Eq. (\ref{Gel}) to the current, we have to 
solve Eqs. (\ref{f1s}) and (\ref{f1ds}) 
for the linear, in 
the electric field, correction to the order parameter
and determine the functions $f_1$ and $f_1^{\dagger}$. As discussed above,
the functions $g_1$ and $\bar g_1$ can be replaced by 
their normal state values to the order to which we work.
To use the 
operator formalism we need 
to evaluate the effect of acting with the
differential operator,
$\Bigl[2\widetilde\Omega_n + {\bf v}\bigl(\nabla-2ie{\cal A}\bigr)\Bigr]^{-1}$,
on the  unperturbed function $f$.
In 
the clean limit 
\begin{equation}
f=2i g\Bigl[2\widetilde\omega_n + {\bf v}\bigl(\nabla-2ie{\cal A}\bigr)\Bigr]^{-1}
\Delta.
\end{equation}
Then the two differential operators can be separated 
\begin{eqnarray}
&&
\Bigl[2\widetilde\Omega_n + {\bf v}\bigl(\nabla-2ie{\cal A}\bigr)\Bigr]^{-1}
\Bigl[2\widetilde\omega_n + {\bf v}\bigl(\nabla-2ie{\cal A}\bigr)\Bigr]^{-1}
\\
\nonumber
&&
\qquad ={1\over 2}(\widetilde\omega_n-\widetilde\Omega_n)^{-1}
\biggl(
\Bigl[2\widetilde\Omega_n + {\bf v}\bigl(\nabla-2ie{\cal A}\bigr)\Bigr]^{-1}
-\Bigl[2\widetilde\omega_n + {\bf v}\bigl(\nabla-2ie{\cal A}\bigr)\Bigr]^{-1}
\biggr),
\end{eqnarray}
and Eq. (\ref{f1s}) 
becomes
\begin{eqnarray}
\label{f1smod}
f_1&=&{e {\bf v A}(f-f(-)) \over i\widetilde\omega_0} 
	+ i(g+g(-))
\Bigl[ 2 \widetilde \Omega_n + {\bf v} (\nabla - 2 i e {\cal A})\Bigr]^{-1}
\Delta_1
\\
\nonumber
&& +i (2\tau)^{-1}(g+g(-))
\Bigl[ 2 \widetilde \Omega_n + {\bf v} (\nabla - 2 i e {\cal A})\Bigr]^{-1}
\langle f_1 \rangle
\\
\nonumber
&&
+ i (2\tau)^{-1}{e{\bf v A} (g-g(-)) \over i\widetilde\omega_0}
\Bigl[ 2 \widetilde \Omega_n + {\bf v} (\nabla - 2 i e {\cal A})\Bigr]^{-1}
\Bigl(\langle f \rangle + \langle f(-)\rangle\Bigl).
\end{eqnarray}
Since
the self consistency condition requires that
\begin{equation}
\label{helpC}
\Delta_1=\pi {\rm g} N(0)T \sum_{\omega_n}\langle f_1\rangle,
\end{equation}
and Eq. (\ref{f1smod}) is a standard Fredholm type integral equation,
it is clear that, since the  function $f$ given in Eq.
(\ref{f}) is a Fourier series in $\phi$, the angular
average of the product $f\cos\phi$ projects out only the component 
proportional to
$\exp(i\phi)$,  the first excited
mode of the order parameter $|m=1\rangle$. This implies that
the linear, in the electric field, change in the order parameter
involves only the first excited state, and has the form
$\Delta_1=C|1\rangle$,
where $C$ is to be determined from Eq. (\ref{f1smod}),
similarly,
$\Delta_1^{\star}=\bar C \langle 1|$.
This result, which was anticipated in
Section \ref{sec:ssolution}, is in agreement with
that of Caroli and Maki \cite{maki1}.
The contribution  to the response due to the dynamical fluctuations of the
order parameter is given by the
second term in Eq. (\ref{Gel}).
Using the functions $f$ and $f^{\dagger}$ from Eqs.
(\ref{f1s}) and (\ref{f1ds}) and the orthogonality
condition given in Eq. (\ref{ortho}), we find that
\begin{eqnarray}
\label{D1fdag}
\overline{\Delta_1 f^{\dagger}}&=&
-{\sqrt{2\pi}\Lambda\Delta^2\over v\sin\theta}CgW'e^{-i\phi}
\\
\overline{\Delta_1^{\star}f}&=&
{\sqrt{2\pi}\Lambda\Delta^2\over v\sin\theta}\bar CgW'e^{i\phi}.
\end{eqnarray}
and the  contribution to the longitudinal and transverse 
electrical current is
\begin{eqnarray}
\label{jfl1}
&&{\bf j}^{fl}_x={\pi\over 4}\sqrt{2\pi} e N(0) \Bigl(\bar C-C\Bigr)
			\Lambda\Delta^2
\int_0^\pi d\theta \sin\theta 
T\sum_{\omega_n} {gW'-g(-)W'(-)\over i\widetilde\omega_0}.
\\
\label{jfly}
&&{\bf j}^{fl}_y=i{\pi\over 4}\sqrt{2\pi} e N(0) \Bigl(\bar C+C\Bigr)
			\Lambda\Delta^2
\int_0^\pi d\theta \sin\theta 
T\sum_{\omega_n} {gW'-g(-)W'(-)\over i\widetilde\omega_0}.
\end{eqnarray}
We see that the ``odd'' part of the dynamical fluctuations of the order
parameter gives rise to a contribution to the longitudinal resistivity,
while the ``even'' part contributes to the Hall current. 
Both of these contributions are proportional to the ``vertex'' function
\begin{equation}
\label{vertex}
V(\omega_0) =T\sum_{\omega_n} {gW'-g(-)W'(-)\over i\widetilde\omega_0},
\end{equation}
describing the coupling between the electric field and the 
excited mode of the order parameter.

Eq. (\ref{f1smod}) is solved in Appendix \ref{sec:C},  
we find that the last
two terms  
result in  small, in $(\Lambda/l)$, contributions, 
and the amplitude of the fluctuations $C$  is given by
\begin{eqnarray}
\label{C}
&&\qquad C={\sqrt2\over 4} ev A\biggl[T\sum_n \int_0^\pi\sin\theta d\theta
{gW'-g(-)W'(-) \over i\widetilde\omega_0}\biggr]
\\
\nonumber
&&\times
\biggl[T\sum_n \int_0^\pi d\theta 
\Bigl[ig{\rm sgn}(\omega_n)W(u_n)-{i\over 2}(g+g(-))
{\rm sgn}(\Omega_n)\Bigl(W(U_n)+{1\over 2}W''(U_n)\Bigr)\Bigr]\biggr]^{-1}
\end{eqnarray}
The denominator on the right hand side of 
Eq. (\ref{C}) is the propagator of the first excited mode
of the order parameter.
In general, the zeroes of this propagator correspond to
the spectrum of propagating modes of the order parameter.
In our case the transverse perturbation due to
the vector potential of the electric field  
couples to the first excited mode of the
order parameter, which is damped, i.e.  there is a finite energy gap
in the spectrum of these excitations at zero frequency.
The response to
a scalar potential is quite different,
there is a propagating mode at zero frequency
\cite{maki1}.  
Since the 
dynamical fluctuations of the order parameter are driven by the
electric field,
 the coupling to the excited mode in the numerator
of Eq. (\ref{C}) is also proportional to
the vertex function 
defined in Eq. (\ref{vertex}).

Evaluating the sums in Eq. (\ref{C})
we find (Eq. (\ref{Cvalue1}))  
\begin{equation}
\label{Cvalues}
C=-\bar  C ={ie\Lambda A \sqrt2\over 1-i\bar\omega\tau}, 
\end{equation}
and therefore there is no contribution to the transverse current due to
the fluctuation term, as expected.
The contribution to the longitudinal current can be evaluated
from Eq. (\ref{jfl1}), it is
\begin{equation}
\label{jfl2}
{\bf j}_{fl}=-i\sqrt\pi N(0)e^2 v^2 \tau 
           \Bigl({\Lambda\Delta\over v}\Bigr)^2
		\int_0^\pi \sin\theta 
		W'\Bigl({i\Lambda\over l\sin\theta}\Bigr) \ {\bf E},
\end{equation}
and the  contribution of the dynamical fluctuations of
the order parameter to the longitudinal conductivity is
\begin{equation}
\label{sxxfl}
\sigma_{xx}^{fl}=4N(0)e^2 v^2 \tau\Bigl({\Lambda\Delta\over v}\Bigr)^2
=12\sigma_n \Bigl({\Lambda\Delta\over v}\Bigr)^2.
\end{equation}

\subsection{Thompson Contribution}
\label{sec:thompsonxx}

We now consider the remaining terms in Eq. (\ref{Gel}). To evaluate their
contribution to the longitudinal current we need the explicit
expressions for the angular averages of the unperturbed anomalous
Green's function $f$ and $f^{\dagger}$, and the linear, in the
electric field,  corrections  $f_1$ and $f_1^{\dagger}$ to the 
distribution function, these are obtained from Eq. (\ref{f1smod})
and its daggered counterpart.

Only one of these terms, the term involving
the angular average of the functions $f_1$ and $f_1^{\dagger}$,
 gives a contribution  to the conductivity at the order
considered here.
A typical term is given by
\begin{eqnarray}
\label{thomhelp}
{\overline{\langle f_1\rangle f^{\dagger}}-
\overline{\langle f_1\rangle f^{\dagger}(-)}
\over 2i\widetilde\omega_0\tau}
&=&i\pi evA \Bigl({\Lambda\Delta\over v}\Bigr)^2 {e^{-i\phi}\over \sin\theta}
{gW'-g(-)W'(-)\over \widetilde\omega_0\tau}
\\
\nonumber
&&\times 
\int_0^\pi {d\theta'  \over 4} \Biggl\{
{gW'-g(-)W'(-) \over i\widetilde\omega_0} \sin\theta'
-2{\Lambda\over v}{g+g(-)\over 1-i\bar\omega\tau}
{\rm sgn}(\Omega_n)\Bigl(W(U_n)+{1\over 2}W''(U_n)\Bigr)\Biggr\},
\end{eqnarray}
where the functions under the integral depend on the angle $\theta'$,
and 
\begin{equation}
U_n={2i\Lambda\widetilde\Omega_n {\rm sgn}(\Omega_n)
\over v\sin\theta}
\end{equation}
in analogy to Eq. (\ref{un}).
 As there is an additional
factor of the scattering time in the denominator, it might be expected
that this contribution is small. However, in the intermediate
region
the renormalized frequency 
$\widetilde\omega_0\sim 1/\tau$ and, since $g$ and $g(-)$
have opposite signs, the contribution of $gW'-g(-)W'(-)$ is of order
one. Therefore the first term contributes to the conductivity at 
the same order as the corrections found previously.
On the other hand, as $g+g(-)=0$ in the intermediate region,
the second term does not contribute to the current.
In the outside region both terms give contributions
to  order $(\Lambda/l)$ which can be neglected.
The contribution to the current 
from Eq. (\ref{thomhelp}) and the corresponding term involving
$\langle f_1^{\dagger}\rangle f$ (which is obtained from equation
(\ref{thomhelp}) by replacing $e^{-i\phi}$ with $e^{i\phi}$) is
\begin{equation}
{\bf j}_{Th1}={\pi^2\over 4} N(0)  e^2 v^2  {\bf A} 
\Bigl({\Lambda\Delta\over v}\Bigr)^2
{1\over 4 i\widetilde\omega_0\tau}
{1\over i\widetilde\omega_0}
\sum_{\omega_n>0}^{\omega_0}
\Biggl[\int_0^\pi\sin\theta d\theta\ \Bigl\{W'+W'(-)\Bigr\}^2\Biggr].
\end{equation}
The Thompson-like contribution to the conductivity is given by
\begin{equation}
\label{sxxthom}
\sigma_{xx}^{Th}=-2N(0)  e^2 v^2 \tau\Bigl({\Lambda\Delta\over v}\Bigr)^2
=-6\sigma_n\Bigl({\Lambda\Delta\over v}\Bigr)^2.
\end{equation}

 In his original
work  Thompson \cite{thompson}  found that there is a 
contribution to the conductivity in the dirty ($l\ll \xi_0$) 
limit due to scattering of quasiparticles
by the dynamical fluctuations of the order
parameter. The main contribution in the dirty limit arose from the outside
region; the contribution of the intermediate region was
smaller by a factor $(l/\xi_0)$. The result obtained here 
for the clean limit is consistent with this picture.
The term contributing to  leading order
is proportional to the angular
average of $f_1$ and exists only in the presence of
 the excited mode of the order parameter as it depends on the angular
average of $f_1$. 
As expected
when  $(l/\xi_0)\gg 1$,
 the relevant contribution comes from the intermediate
region. In the presence of a transport current the vortex lattice
moves, and individual vortices are deformed. As a result, 
additional scattering 
of quasiparticles by 
the vortices  gives rise to a negative
contribution to the conductivity given in Eq. (\ref{sxxthom}).

\subsection{Longitudinal Conductivity}

The longitudinal conductivity of a clean type-II superconductor in the mixed
state is obtained by combining the  results for
the quasiparticle current from Eqs. (\ref{sxxqp1}) and
(\ref{sxxqptau}), the current due to the dynamical fluctuations
of the order parameter from Eq. (\ref{sxxfl}) and
the current due to the Thompson terms from Eq. (\ref{sxxthom}).
We notice that reduction in the quasiparticle contribution
to the conductivity due to additional scattering off 
the ground state of the vortex lattice
(Eq. (\ref{sxxqp1}))
 and  the excited modes of the order parameter
(Thompson terms)
is compensated to order 
$(\Lambda\Delta/v)^2$ by the increase in the current due to 
dynamical fluctuations of the order parameter
The conductivity then is  given by  Eq. (\ref{sxxqptau})
\begin{equation}
\label{sxx}
\sigma_{xx}={1\over 3}N(0)e^2v^2\tau_{eff}=
\sigma_n \biggl[ 1+2
                \biggl({\Lambda \Delta \over v}\biggr)^2
               \biggl(\ln\biggl({2 v^2 \over\Lambda^2\Delta^2}\biggr)-1
			\biggr) \biggr],
\end{equation}
 that is the modification of the longitudinal conductivity
upon entering the superconducting state is determined solely by
the increase in the effective mean free path due to the
suppression of the density of states at the Fermi level as the 
superconducting gap opens. The increase 
in the mean free path is a non-analytic function
of the order parameter.

\section{Hall Effect}
\label{chap:sxy}

\subsection{Stability of the Leading Order Solution}
\label{sec:stable}

In determining the density of states
and the longitudinal conductivity we have neglected terms of the order of 
cyclotron frequency not only in the  linearized quasiclassical equations, 
but also in the leading order
equations 
(\ref{eqg0})-(\ref{eqfd}). To investigate the 
behavior of the transverse 
conductivity the gradient terms in Eq. (\ref{g0sup})
have to be taken into account, and the 
solution for the propagator $\widehat g$  at zeroth order
in the electric field has to be obtained
to order $\omega_c$. Instead of attempting to solve 
equation  (\ref{g0sup}) in  full, 
we will show here that the solution obtained in
Section \ref{chap:dos} is still valid when terms of order of the
cyclotron frequency are included in the equations. 

We saw in Section \ref{chap:eqns}  that, as the matrix combination
$\widehat\sigma_z\widehat g + \widehat g\widehat\sigma_z$ has only 
diagonal elements,
 the Lorentz force acts
only on the quasiparticle (diagonal) part of the propagator. 
The function $g$ given in Eq. (\ref{gs})
does not depend on the azimuthal angle $\phi$, and, therefore,
there is no correction to this function from the Lorentz force term. 
Next we observe that the term involving the gradient of the
propagator in the third line of Eq. (\ref{g0sup}) is
proportional to the same combination of matrices as the Lorentz term.
Since in the BPT approximation  the function $g$ is replaced by its
spatial average, this term vanishes.
For an s-wave superconductor the order parameter $\Delta$ is
constant at any point at the Fermi surface, and its 
derivative with respect to the components of momentum parallel
to the Fermi surface vanishes, which means that the last term in the fourth
line of equation  (\ref{g0sup}) can be ignored. The momentum derivative
of the self energy due to impurity scattering vanishes for the
same reason.

We now consider the remaining terms in Eq. (\ref{g0sup}). 
Omitting the subscript since in this Section 
we only consider functions at leading order, we write 
the first of these terms in the matrix form
\begin{equation}
\widehat M=
{i \over 2}\biggl[ {\partial\widehat\Delta\over\partial {\bf R}}
{\partial \widehat g \over\partial{\bf p}_{\|}}
+{\partial \widehat g \over\partial {\bf p}_{\|}}
{\partial\widehat\Delta\over\partial {\bf R}}\biggr].
\end{equation}
The off diagonal elements of this matrix are proportionl to the
trace of the quasiclassical propagator and vanish in accordance with
the normalization
condition.
The contribution from the term $\widehat M$ to the
equation for the quasiparticle part of the distribution function is
\begin{equation}
\label{m}
M_{11}=
{i \over 2}\biggl[{\partial\Delta\over\partial {\bf R}}
{\partial f^{\dagger}\over\partial{\bf p}_{\|}}
+{\partial\Delta^{\star}\over\partial {\bf R}}
{\partial f\over\partial{\bf p}_{\|}}\biggr].
\end{equation}
To spatially average this term and determine its contribution to
the diagonal part of the propagator we need to recast
the gradient operators in terms of 
raising and lowering operators $a$ and $a^{\dagger}$
and the azimuthal and polar angles $\phi$ and $\theta$.
We find 
\begin{equation}
\label{hallop}
{\widehat\partial\over\partial {\bf R}}{\partial \over\partial{\bf p}_{\|}}
={1\over \sqrt2 p\Lambda}
\biggl[\Bigl(a e^{-i\phi}-a^{\dagger} e^{i\phi}\Bigr)\cos\theta
				{\partial\over\partial\theta}
-{i\over\sin\theta}\Bigl(a e^{-i\phi}+a^{\dagger} e^{i\phi}\Bigr)
				{\partial\over\partial\phi}\biggr]
\end{equation}
Here the hat denotes the gauge invariant gradient
\begin{equation}
{\widehat\partial\over\partial {\bf R}}=
{\partial\over\partial {\bf R}}\pm 2ie{\cal A},
\end{equation}
and the operator with the plus sign acts on $\Delta^{\star}$ while
the operator with the minus sign acts on $\Delta$. A direct check
using the solution obtained in Section \ref{chap:dos} shows that 
the terms breaking gauge invariance  vanish after spatial averaging, 
consequently,  the
gradient can be replaced by its gauge invariant counterpart, as
expected for an operator acting on the order parameter. For the ground state
of the vortex lattice a direct check shows that 
this term  does not result in any correction to
the unperturbed propagator.
The contribution  from the term involving 
the spatial derivative of the self energy  vanishes in complete
analogy to the term just discussed as their structure is identical.

Therefore the solution 
of the quasiclassical equations obtained in Section 
\ref{chap:dos}
also satisfies the quasiclassical equations 
when  terms of the 
order of cyclotron frequency are taken into account.

\subsection{Linearized Equations for the Transverse Response}
\label{sec:halleq}

We now consider the linearized Eq. (\ref{g1sup}). In the regime
when $\omega_c\tau \ll 1$ terms of the order of cyclotron frequency 
can be included in the calculation of the response
function  perturbatively. We therefore solve
for the linear, in the cyclotron frequency, corrections to 
the averaged response function $g_e=g_1-\bar g_1$ obtained in the calculation
of the longitudinal conductivity in the preceding section. 

Since the Hall conductivity in the normal state is proportional
to the square of the scattering time $\tau$, we can expect that
the most relevant contributions to the transverse conductivity in the
vortex state are also proportional to $\tau^2$, other 
contributions to the Hall effect 
are smaller by a factor $(\Lambda /l)$. Therefore we
keep in the equations only terms that contribute to this order.
If now $\delta\widehat g$ is the part of the propagator linear
in the cyclotron frequency, we arrive at the following 
equation for the function $\delta g_e=\delta g_1 - \delta \bar g_1$
\begin{eqnarray}
\label{ghall}
\delta g_e&=& {i\omega_c\over i\widetilde\omega_0}
{\partial\over\partial\phi}\Bigl(g_1-\bar g_1\Bigr)
+{1\over i\widetilde\omega_0}
	\biggl\{\delta\Delta_1^{\star}\Bigl(f-f(-)\Bigr)+
		\delta\Delta_1\Bigl(f^{\dagger}-f^{\dagger}(-)\Bigr)\biggr\}
\\
\nonumber
&&
+{1\over 2i \widetilde\omega_0\tau} \biggl\{\bigl( 
 \langle \delta f_1^{\dagger} \rangle f-
 \langle\delta f_1^{\dagger} \rangle f(-)\bigr) + 
 \bigl(\langle \delta f_1 \rangle f^{\dagger}- 
 \langle \delta f_1 \rangle f^{\dagger}(-)\bigr)
					\biggr\}
\\
\nonumber
&&
-{i\over 2i\widetilde\omega_0}\biggl\{
{\partial \Delta_1^{\star} \over \partial {\bf R}}
		    {\partial \over \partial {\bf p}_{||}}
		   \bigl(f+f(-)\bigr)
                 +{\partial \Delta_1 \over \partial {\bf R}}
		  {\partial \over \partial {\bf p}_{||}}
                   \bigl(f^{\dagger}+f^{\dagger}(-)\bigr) 
+2{\partial \Delta^{\star} \over \partial {\bf R}}
		    {\partial f_1^{\dagger} \over \partial {\bf p}_{||}}
		+2{\partial \Delta \over \partial {\bf R}}
		{\partial f_1	 \over \partial {\bf p}_{||}}	
    \biggr\}.
\end{eqnarray}
Here $g_1-\bar g_1$ is given by Eq. (\ref{Gel}). There are now two distinct
contributions both to the term involving the fluctuation of the order
parameter and to the Thompson term. One  reason these terms contribute to the
transverse response is that they give rise to additional
scattering due to dynamical fluctuations of the order
parameter induced by the electric field, as we saw in
the previous section. 
When  the quasiparticle  trajectories are bent by
the magnetic field, this additional scattering renormalizes the
Hall conductivity. This effect is contained in
the first term in Eq. (\ref{ghall}), since the function
$g_1-\bar g_1$ contains the contributions of the Thompson terms
and the fluctuations of the order parameter induced by the 
electric field. The other contribution to  
these terms is due to 
 fluctuations of the order parameter
induced by the Lorentz force, these fluctuations result in 
corrections to the transverse conductivity and are contained in
the terms involving $\delta\Delta_1$ and $\delta f_1$ in 
Eq. (\ref{ghall}).
Finally, the terms involving the gradient of the order parameter
 contribute to  order $\tau^2$ for the same
reason that the Thompson term contributes to the longitudinal
conductivity, namely, that there is an additional factor of
the scattering time $\tau$ in the amplitude of the order parameter
fluctuations $C$ 
and in the functions $f_1$ and $f_1^{\dagger}$,
so that the overall contribution is of order $\tau^2$. 
 This anomalous contribution
to the transverse conductivity arises because the gradients
of the order parameter created by the moving and deformed
vortex lattice act as driving forces (analogous to the Magnus force) 
in the transport-like equations.
The remaining terms
in Eq. (\ref{g1sup}) contribute at higher order
in $\Bigl(\Lambda/l\Bigr)$.

The equations for the corrections to the off diagonal elements of
the matrix distribution function are
\begin{eqnarray}
\label{fhall}
\bigl[ 2 \widetilde \Omega_n + {\bf v}_f (\nabla - 2 i e {\cal A})\bigr] 
\delta f  & =&
-{e\over 2m}{\bf A}\Bigl(\nabla -2ie{\cal A}\Bigr)\Bigl(f-f(-)\Bigr)
+i\Delta \delta g_e + i\delta\Delta_1\Bigl(g+g(-)\Bigr)
\\
\nonumber
&& +{i\over 2}{\partial \Delta \over \partial {\bf R}}
		 {\partial \over \partial {\bf p}_{||}}
		\Bigl(g_1+\bar g_1\Bigr)
+{i\over 2}{\partial \Delta_1\over \partial {\bf R}}
	 {\partial \over \partial {\bf p}_{||}}
		\Bigl(g+g(-)\Bigr)
\end{eqnarray}
and
\begin{eqnarray}
\label{fdhall}
{\rm   }  
\bigl[ 2 \widetilde \Omega_n - {\bf v}_f (\nabla + 2 i e {\cal A})\bigr] 
\delta f^{\dagger}  & =&
-{e\over 2m}{\bf A}\Bigl(\nabla +2ie{\cal A}\Bigr)
	\Bigl(f^{\dagger}-f^{\dagger}(-)\Bigr)
+i\Delta^{\star}\delta g_e + i\delta\Delta_1^{\star}\Bigl(g+g(-)\Bigr)
\\
\nonumber
&& +{i\over 2}{\partial \Delta^{\star}\over \partial {\bf R}}
		 {\partial \over \partial {\bf p}_{||}}
		\Bigl(g_1+\bar g_1\Bigr)
+{i\over 2}{\partial \Delta_1^{\star}\over \partial {\bf R}}
	 {\partial \over \partial {\bf p}_{||}}
		\Bigl(g+g(-)\Bigr).
\end{eqnarray}
 In the normal state
there is no angular dependence to the unperturbed function $g$, and
also
$g_1+\bar g_1=0$, so that the terms in the last line of each equation vanish.
We can now solve Eqs. (\ref{fhall}) and 
(\ref{fdhall}) to determine the contributions to the 
dynamical fluctuations of the order parameter induced by  the
Lorentz force. We can then evaluate  the contributions to the
transverse conductivity term by term.

\subsection{Hall Conductivity}

The quasiparticle part of the response function is
\begin{equation}
\delta g_e^{qp}= - 2 e v A \sin\theta\sin\phi 
{i\omega_c\over i\widetilde\omega_0}
	 {(g - g(-))\over i \widetilde\omega_0}.
\end{equation}
The contribution to the conductivity from the intermediate frequency range
 is readily evaluated; 
the correction to the transverse conductivity due to 
additional scattering off the vortex lattice
\begin{equation}
\label{sxyqpdos}
\Delta\sigma_{xy}^{qp1}=-6\sigma_n \omega_c \tau (\Lambda \Delta / v)^2
\end{equation}
and the contribution to the Hall conductivity due to the
modification of the scattering time
\begin{equation}
\label{sxyqptau}
\sigma_{xy}^{qp2}={1\over 3}N(0)e^2v^2 \tau_{eff} (\omega_c \tau_{eff})=
\sigma_n \omega_c \tau \biggl[ 1+ 4
                \biggl({\Lambda \Delta \over v}\biggr)^2
               \biggl(\ln\biggl({2 v^2 \over\Lambda^2\Delta^2}\biggr)-1
			\biggr) \biggr].
\end{equation}
In addition
there is a quasiparticle  contribution to the Hall conductivity from
the outer frequency range which is formally divergent
\begin{equation}
\label{jcancel}
{\bf j}_y^{an1}=- {1\over 4}N(0)e^2 v^2A {i\omega_c\over\bar\omega}
\int_0^\pi \sin^3\theta d\theta \biggl[ (P-1) + 
\Bigl({\Lambda\bar\omega\over v\sin\theta}\Bigr)
 	{\partial P\over\partial\omega}\biggr].
\end{equation}
Since $P'$ evaluated at zero frequency 
is purely imaginary, the second term describes a small
correction to the Meissner-like term.
The first term in this equation, on the other hand, has no physical
meaning
 and must disappear from the final expression for the current. 

It is in fact cancelled by the contribution of the fluctuations of the
order parameter 
\begin{equation}
\label{jflhall1}
{\bf j}_y^{fl}=2i\pi e  \omega_c CN(0) {\sqrt{2\pi}\Lambda\Delta^2\over v}
\int d^2 s {\bf v}{\sin\phi\over\sin\theta}
T\sum_{\omega_n} {gW'-g(-)W'(-)\over (i\widetilde\omega_0)^2}
\end{equation}
from the outer range, where the first term in the expansion of
the vertex function 
\begin{equation}
T\sum_{out} {gW'-g(-)W'(-)\over (i\widetilde\omega_0)^2}=
-{2i\over\bar\omega}T\sum_{\omega_n>0}{\partial W'\over\partial\omega}+\ldots
\end{equation}
is formally divergent. The remaining contribution from the outer range 
is obtained by expanding the coefficients $C$ and $\bar C$ in small
quantity $\bar\omega\tau$, it is
\begin{equation}
\label{jflhallout}
{\bf j}_y^{fl2}=6 \sigma_n (\omega_c\tau) 
\Bigl({\Lambda\Delta\over v}\Bigr)^2 E.
\end{equation}
This contribution is cancelled by that of the intermediate frequency range
in Eq. (\ref{jflhall1}), so that there is no net contribution
to the transverse conductivity due to the dynamical fluctuations of the 
order parameter driven by the electric field.
This result is consistent with  the predictions of time dependent
Ginzburg-Landau theory\cite{Dorsey}. The terms 
considered so far in this Section correspond directly to those 
contained in TDGL, which is an effective theory treating only
the fluctuations of the order parameter, while the quasiparticle
contribution is taken to be at the normal state value.
In the TDGL approach
the Lorentz force has no effect
on the dynamics of the order parameter, and
there is no correction $\delta \Delta_1$ due to this force.

In the present analysis, however, 
the equations for the quasiparticle propagators and  the amplitude of the
order parameter fluctuations are coupled, so that even though the
Lorentz force does not appear explicitly in the equation for
$\delta f$, it 
introduces changes in the diagonal part 
of the distribution function $g_1-\bar g_1$ and therefore
brings about further modification 
$\delta \Delta_1$ of the order parameter.
To find this contribution
we have to solve Eqs. (\ref{fhall})
and (\ref{fdhall}) for the changes in the order parameter
$\delta\Delta_1=\delta C | 1\rangle$ and 
$\delta\Delta_1^{\star}=\delta \bar C \langle 1 |$.
The solution, given in Appendix \ref{sec:C},
 follows the same steps as in the calculation of 
the longitudinal conductivity,
We find that only the term involving $\delta g_e$ contributes at the order
to which we work, and
\begin{equation}
\label{dC}
\delta C= \delta \bar C=ie\Lambda A \sqrt2 (\bar\omega\tau)(\omega_c\tau).
\end{equation}
and, since we saw in Eq. (\ref{jfly})
that
the ``even'' part of  dynamical fluctuations of the
order parameter contributes to the transverse part of the conductivity
\begin{equation}
\label{sxyfl}
\sigma_{xy}^{fl}= 6 \sigma_n (\omega_c\tau) \Bigl({\Lambda\Delta\over v}\Bigr)^2. 
\end{equation}
The dynamical fluctuations of the order parameter
driven by the Lorentz force tend to increase the transverse conductivity.

Similarly, there are two 
parts to the Thompson terms: one is due to 
 the longitudinal response, while the other is
due to the linear in  the Lorentz force 
corrections to the off diagonal distribution functions. 
The contribution
to the transverse conductivity due to the first of the Thompson terms
is 
the longitudinal contribution multiplied by  $\omega_c\tau$,
\begin{equation}
\label{sxyth1}
\sigma_{xy}^{Th1}=-6 \sigma_n (\omega_c\tau)\Bigl({\Lambda\Delta\over v}\Bigr)^2.
\end{equation}
To determine the contribution to the current from the second
Thompson term, we use the angular
 averages $\langle \delta f\rangle$ and
$\langle \delta f^{\dagger}\rangle$ given in Appendix \ref{sec:C},
to find that its contribution doubles that given
in Eq. (\ref{sxyth1}), 
 so that the total contribution of the Thompson terms to
the transverse conductivity is
\begin{equation}
\label{sxythom}
\sigma_{xy}^{Th}=-12 \sigma_n (\omega_c\tau)\Bigl({\Lambda\Delta\over v}\Bigr)^2.
\end{equation}
In addition to the
scattering of  quasiparticles 
by  the fluctuations of the order parameter induced
by the applied electric field, which tends to reduce
the current, quasiparticles also undergo
additional scattering off the dynamical fluctuations driven
by the Lorentz force, which again tends to reduce the transverse
response. In this sense, the Lorentz force results in anisotropic scattering
of the quasiparticles by the fluctuations of the order parameter.

The last two terms, the gradient terms
 in Eq. (\ref{ghall}) have  a structure
identical to  that of the
terms considred in  Section \ref{sec:stable}. Their contribution to the
current can be evaluated using the operator approach as shown in 
Appendix \ref{sec:operators}.
First, consider the  term involving the gradient
of the dynamical fluctuations of the order parameter $\Delta_1$ and
$\Delta_1^{\star}$. These fluctuations involve the first excited mode
of the order parameter, which,
when they are acted upon by the  annihilation and creation operators
in Eq. (\ref{gradop}), gives terms proportional to
 the ground state and the second
excited state of the order parameter, respectively. Then spatial averaging
projects out the same modes from the functions $f$ and $f^{\dagger}$
given in Eqs. (\ref{f}) and (\ref{fdagger}).
In the second of these terms the gradient operator acts on the ground state
of the vortex lattice, $\Delta$, which the creation operator promotes to the
first excited state. Spatial averaging now projects out the
first excited component from the   functions $f_1$ and $f_1^{\dagger}$.
The   contribution to the transverse current due to these terms, 
is found to be
\begin{equation}
\label{jhallgr}
{\bf j}_y^{gr}=-3\sigma_n {\Delta^2\tau\over \epsilon_f} E.
\end{equation}
Since
\begin{equation}
{\Delta^2\tau\over \epsilon_f}={2\Delta^2\tau\over mv^2}=
2 \Bigl({\Lambda\Delta\over v}\Bigr)^2 {\tau\over m\Lambda^2}=
4\omega_c\tau \Bigl({\Lambda\Delta\over v}\Bigr)^2,
\end{equation}
 the  contribution to the transverse conductivity due to these
terms is given by
\begin{equation}
\label{sxygr}
\sigma_{xy}^{gr}=12\sigma_n \omega_c\tau \Bigl({\Lambda\Delta\over v}\Bigr)^2.
\end{equation}
The induced gradients of the order parameter
enhance the transverse conductivity.

The total  transverse conductivity
is the sum of all the contributions considered here.
We find
that the modification of the quasiparticle Hall current
due to additional scattering off the vortex lattice
given in Eq. (\ref{sxyqpdos})
is exactly compensated by the 
enhancement 
of the transverse current due to Lorentz force driven fluctuations 
of the order parameter obtained in Eq. (\ref{sxyfl})
The Thompson contribution due to 
 additional scattering  by the deformed and
moving vortex lattice is given in
Eq. (\ref{sxythom}) and is cancelled by 
the enhancement of  the
transverse conductivity due to the forces generated by 
gradient of the excited mode of 
the order
parameter found in Eq. (\ref{sxygr}).
As a result, the behavior of the transverse conductivity $\sigma_{xy}$ is
determined solely by the  modification 
of  the effective elastic scattering time $\tau_{eff}$
and is given by equation
(\ref{sxyqptau})
\begin{equation}
\label{sxy}
\sigma_{xy}^{qp2}={1\over 3}N(0)e^2v^2 \tau_{eff} (\omega_c \tau_{eff})=
\sigma_n \omega_c \tau \biggl[ 1+ 4
                \biggl({\Lambda \Delta \over v}\biggr)^2
               \biggl(\ln\biggl({2 v^2 \over\Lambda^2\Delta^2}\biggr)-1
			\biggr) \biggr].
\end{equation}
For the dc conductivity this change is due to the decrease in the number of
states at the Fermi surface available for scattering as the 
superconducting gap opens.

\section{Conclusions and Discussion}

We now plot qualitatively the longitudinal resistivity (Figure \ref{fig:rxx}), 
the transverse conductivity (Figure \ref{fig:sxy})
and the Hall angle (Figure \ref{fig:tan})
as functions of the applied magnetic field for
Niobium. The order parameter, which is linear in the applied magnetic field
in the high field regime, is given by the expression due to Maki and
Tsuzuki \cite{makit}
\begin{equation}
\label{d2}
\Delta^2 = { 1 \over  \pi N(0)} {H_{c2} - H \over \beta_A (2 \kappa_2^2 -1)}
           \biggl( H_{c2} - {T \over 2} {d H_{c2} \over dT} \biggr).
\end{equation}
and the values of the superconducting material parameters  were taken from
Refs. \cite{finn,mccon,nbdos}. The longitudinal resistivity in Figure 
\ref{fig:rxx}
has a pronounced increase
in slope as a function of the magnetic field below the
superconducting transition due to the logarithmic dependence
in equation (\ref{sxx}). The transverse conductivity shown in
Figure \ref{fig:sxy} is
enhanced below the upper critical field and has
negative curvature in the high field region. 
The negative curvature
arises from the competition between the
enhancement due to the  increase in the effective mean free path 
and the linear decrease of the cyclotron frequency with the field;
the Hall conductivity is substantially enhanced when compared to
the linear decrease expected from the normal state behavior.
While the transverse conductivity is proportional to the 
square of the scattering 
time, the  Hall angle 
\begin{equation}
\tan \theta_H=\sigma_{xy}/\sigma_{xx}=\omega_c\tau_{eff}
\end{equation} 
is only linearly dependent on the scattering time and the corresponding
nonlinear dependence on magnetic field is weaker, as can be seen in
Figure \ref{fig:tan}.
Finally, as the transverse resistivity 
\begin{equation}
\rho_{xy}={\sigma_{xy}\over\sigma^2_{xx}+\sigma^2_{xy}}
\approx \sigma_{xy}/ \sigma_{xx}^2
\end{equation}
 is independent of the effective scattering time, 
 it  remains 
linear in magnetic field upon entering the superconducting state
 with the same slope as in the normal metal.  
This behavior is to be contrasted with
that of Bardeen-Stephen model \cite{BS}, where the resistivity is modified
 and is linear in the magnetic field,
but the Hall angle obeys the same linear law as in the normal state. 
The Nozieres-Vinen theory \cite{NV}, on the other hand,  
which predicts that the Hall angle should be constant in the flux-flow regime
below $H_{c2}$ at variance with the result of this work,
also finds that the transverse resistivity is identical to that
of the normal state, although the individual components
of the conductivity tensor are quite  different from those found here.

A comparison can be made with the experimental data of Fiory and Serin
\cite{fiory} on high purity Nb. These experiments find
a transverse resistivity in the 
flux-flow regime which is linear in the applied magnetic field 
over a wide range of fields below $H_{c2}$.
The Hall angle, however, flattens or even increases above its value at
$H_{c2}$ before decreasing at lower fields.
These results are more 
suggestive of the behavior given here than the original interpretation 
given in terms of the Nozieres-Vinen theory. 
Also, the longitudinal resistivity found in Ref. \cite{fiory}
has a distinct increase in slope just below the upper
critical field, which is consistent with the  behavior 
discussed above. Detailed comparisons with the
results of this work  are difficult to make, since the 
authors of Ref.\cite{fiory} 
used a high current density to reduce the pinning effects
and achieve the flux flow regime; as a result, the magnetoresistance
is significant and the longitudinal resistivity in the normal state 
varies with magnetic field.
We find the qualitative agreement with the experiment encouraging and
suggest that
 more experimental work is needed to make a more 
detailed comparison with the 
theory.
To conclude, we have presented here a new approach to the calculation of
the transport coefficients of a clean type-II superconductor in the
vortex state in the high-field regime and used it to determine
the Hall conductivity and the Hall angle of an s-wave superconductor
in this regime. We find that the field dependence of the Hall conductivity
in the high field regime, which is non-analytic, is entirely due to the
change in the density of quasiparticle states at the Fermi level in 
the superconducting state. At the same time we find that the field dependence
of the transverse resistivity below the upper critical field
remains unchanged.

\acknowledgements

One of us (AH) would like to thank A. T. Dorsey, D. Rainer and K. Scharnberg
for important discussions. This research was supported in part by
the National Science Foundation through Grant No. DMR9008239.
%%%%%%%%%%%%%%%%%%%%%%%%%%%%%%%%%%%%%%%%%%%%%%%%%%%%%%%%%%%%%%%%%%

\appendix
\section{The Operator Formalism}
\label{sec:operators}

If $|m\rangle$ is the
$m$-th excited mode of the order parameter we have
\begin{eqnarray}
\label{op1}
&&[2 \widetilde \omega_n +
{\bf v}_F  ( \nabla - 2 i e {\cal A}) ]^{-1} | m \rangle 
=
{\rm sgn}(\omega_n) \int_0^\infty 
                 \exp\biggl(- \bigl(2 \widetilde \omega_n +
{\bf v}_F  ( \nabla - 2 i e {\cal A})
             \bigr)
                       {\rm sgn}(\omega_n) t \biggr)
                         dt  \ | m \rangle 
\\
\nonumber
&&\qquad
={\rm sgn}(\omega_n) \int_0^\infty 
\exp\bigl(-2 \widetilde \omega_n {\rm sgn}(\omega_n)t 
                                       \bigr)
        \exp \biggl( -{v \sin \theta \over \sqrt2 \Lambda}
                      [ a e^{-i \phi} - a^{\dagger} e^{i \phi}]
                         {\rm sgn} (\omega_n) t \biggr)
                            dt  \ | m \rangle.
\end{eqnarray}
We use the operator identity
\begin{equation}
e^{A+B}=e^A e^B e^{-{1 \over 2}[A,B]},
\end{equation}
where $[A,B]=AB-BA$ denotes a commutator, 
to separate the creation and annihilation operators and 
rewrite Eq. (\ref{op1}) as
\begin{eqnarray}
\label{op2}
[2 \widetilde \omega_n +
{\bf v}_F  ( \nabla - 2 i e {\cal A}) ]^{-1} | m \rangle &=&
{\rm sgn}(\omega_n) \int_0^\infty 
            dt \exp\bigl(-2 \widetilde \omega_n {\rm sgn}(\omega_n)t 
                              -{v^2  \sin^2 \theta \over
                                  4 \Lambda^2} t^2 \bigr)
\\
\nonumber
&&
  \times\exp \Bigl[ {vt \sin\theta \over \sqrt2 \Lambda} {\rm sgn}(\omega_n) 
                                           e^{i\phi} a^{\dagger}\Bigr]
      \exp \Bigl[- {vt {\rm sin}\theta \over \sqrt2 \Lambda} 
{\rm sgn}(\omega_n)     
                                        e^{-i\phi} a \Bigr]  | m \rangle.
\end{eqnarray}
We now
write the exponentials as infinite series in powers of the arguments
to find
\begin{eqnarray}
\label{op3}
[2 \widetilde \omega_n +
{\bf v}_F  ( \nabla - 2 i e {\cal A}) ]^{-1} | m \rangle &=&
= \sum_{m_2=0}^{\infty} \sum_{m_1=0}^m \int_0^\infty dt
       \exp\bigl(-2 \widetilde \omega_n {\rm sgn}(\omega_n)t 
                               -{v^2  \sin^2 \theta \over
                                  4 \Lambda^2} t^2 \bigr)
{ (-1)^{m_1} \over m_1! m_2!} e^{i(m_2 -m_1) \phi}
\\
\nonumber
&&
\times
 \bigl({\rm sgn}(\omega_n) \bigr)^{m_1+m_2+1}
\biggl[{vt {\rm sin}\theta \over \sqrt2 \Lambda}\biggr]^{m1+m2}
{(a^{\dagger})}^{m_2} (a)^{m_1} | m \rangle.      
\end{eqnarray}
In the integral, the parameter $t$ can be replaced with a differential
operator 
\begin{equation}
t= \biggl[ - {1 \over 2} {\rm sgn}(\omega_n)
{\partial \over \partial \widetilde\omega_n} \biggr],
\end{equation}
and the integral can be evaluated 
\begin{eqnarray}
&&\qquad\int_0^\infty dt \exp\Bigl[-2 \widetilde \omega_n {\rm sgn}(\omega_n)t 
                               -{v^2  \sin^2 \theta \over
                                  4 \Lambda^2} t^2 \Bigr]
\biggl[{vt {\rm sin}\theta \over \sqrt2 \Lambda}\biggr]^{m1+m2}
\\
\nonumber
&&= \biggl[{v{\rm sin}\theta \over \sqrt2 \Lambda}\biggr]^{m1+m2}
   \biggl[ - {1 \over 2} {\rm sgn}(\omega_n)
{\partial \over \partial \widetilde\omega_n} \biggr]^{m_1+m_2}
\int_0^\infty dt \exp\bigl(-2 \widetilde \omega_n {\rm sgn}(\omega_n)t 
                               -{v^2  \sin^2 \theta \over
                                  4 \Lambda^2} t^2 \bigr)
\\
\nonumber
&&={\sqrt\pi \Lambda \over v \sin\theta} 
\bigl( -{i \over \sqrt2} \bigl)^{m_1+m_2}
 W^{(m_1+m_2)} (u_n),
\end{eqnarray}
where $W(u)=e^{-u^2}{\rm erfc}(-iu)$, $W^{(m)}$ denotes the $m$-th derivative
and
\begin{equation}
u_n= {2 i \widetilde \omega_n \Lambda{\rm sgn}
                                     (\omega_n)
                                     \over
                                       v \sin \theta}.
\end{equation}
The main result is
\begin{equation}
\label{opmain1}
[2 \widetilde \omega_n +
{\bf v}_f  ( \nabla - 2 i e {\cal A}) ]^{-1} | m \rangle 
=
{\sqrt{\pi} \Lambda \over v \sin \theta}
            \sum_{m_2=0}^{\infty} \sum_{m_1=0}^m 
              D_m^{m_1m_2} e^{i(m_2 -m_1) \phi} 
              | m+m_2-m_1 \rangle,
\end{equation}
 where 
\begin{equation}
\label{opd1}
D_m^{m_1m_2}=
        {\sqrt{m!} \sqrt{(m-m_1+m_2)!} \over (m-m_1)! m_1! m_2!}
          (-1)^{m_1}      (-{i \over \sqrt2})^{m_1+m_2}
\bigl({\rm sgn}(\omega_n) \bigr)^{m_1+m_2+1}                  
                        W^{(m_1+m_2)}
                              (u_n).
\end{equation}
We make extensive use of two special cases of Eqn.(\ref{opmain1}):
\begin{equation}
\label{oper0}
( 2 \widetilde \omega_n +
{\bf v} ( \nabla - 2 i e {\cal A}) )^{-1} | 0 \rangle  =
     {\sqrt{\pi} \Lambda \over v \sin \theta}
     \sum_{m=0}^{\infty} {1 \over \sqrt{m!}} (-{i \over \sqrt2})^m e^{i m \phi}
({\rm sgn}(\omega_n))^{m+1}
             W^{(m)} (u_n) | m \rangle
\end{equation}
and
\begin{eqnarray}
\label{oper1}
&& \qquad(2 \widetilde \omega_n +
{\bf v}  ( \nabla - 2 i e {\cal A}) )^{-1} | 1 \rangle  =
                      {\sqrt{\pi} \Lambda \over v \sin \theta}
               \sum_{m=0}^{\infty} 
  {1 \over \sqrt{m!}} (-{i \over \sqrt2})^m ({\rm sgn}(\omega_n))^{m+1}
\\
\nonumber
&&\qquad\qquad\qquad\times 
\biggl[ \sqrt{m+1} e^{i m \phi} W^{(m)}(u_n)| m+1 \rangle+
      ({i \over \sqrt2}) ({\rm sgn}(\omega_n)) e^{i (m-1) \phi} 
	W^{(m+1)}(u_n)| m \rangle  
             \biggr].
\end{eqnarray}
Equations for the daggered quantities are obtained by replacing 
the phase $im\phi$ by its conjugate $-im\phi$, changing the sign
of $(i/\sqrt2)$, and using a bra vector instead of the ket vector.

To determine the quasiclassical Green's 
function $g$ we need the spatial average
$\overline{f f^{\dagger}}$. Using Eq. (\ref{op3}) we have
\begin{eqnarray}
\overline{f f^{\dagger}}&=& \int d^3 {\bf R} f f^{\dagger}
=- 4 g^2 \Delta^2 \int_0^{\infty} dt_1 dt_2 
             \exp\biggl(-2 \widetilde \omega_n {\rm sgn}(\omega_n) (t_1 + t_2) 
                               -{v^2  \sin^2 \theta \over
                                  4 \Lambda^2} (t_1 + t_2)^2 \biggr) 
\\ 
&=& i \sqrt\pi g^2 \biggl( {2 \Lambda \Delta \over v \sin \theta} \biggr)^2
W^{\prime} \biggl({2 i \widetilde \omega_n \Lambda {\rm sgn}(\omega_n)
                                     \over
                                       v \sin \theta}\biggr).
\end{eqnarray}
Eq. (\ref{gs}) obviously follows from the last line. 

In the calculation of the transverse conductivity we will need to
rewrite the gradient operators in terms of 
 creation and annihilation operators.
Since a gauge invariant gradient can be written as
\begin{eqnarray}
\label{gradx}
&&{\widehat\partial \over\partial x}={1\over \sqrt2 \Lambda}
	\Bigl[a-a^{\dagger}\Bigr]
={1\over \sqrt2 \Lambda}\Bigl[b-b^{\dagger}\Bigr]
\\
\label{grady}
&&{\widehat\partial \over\partial y}=-{i\over \sqrt2 \Lambda}
	\Bigl[a+a^{\dagger}\Bigr]
={i\over \sqrt2 \Lambda}\Bigl[b+b^{\dagger}\Bigr],
\end{eqnarray}
and the the momentum gradient in the
direction parallel to the Fermi surface is
\begin{equation}
\label{gradsph}
{\partial \over\partial{\bf p}_{\|}}=
{1\over p}\widehat{\bf e}_\theta {\partial\over\partial\theta}
+{1\over p\sin\theta}\widehat{\bf e}_\phi {\partial\over\partial\phi},
\end{equation}
where $\widehat{\bf e}_\theta$ and $\widehat{\bf e}_\phi$ are the unit vectors
in $\theta$ and $\phi$ direction in the spherical coordinates,
we obtain
\begin{eqnarray}
\label{gradop}
{\widehat\partial\over\partial {\bf R}}{\partial \over\partial{\bf p}_{\|}}
&=&{1\over \sqrt2 p\Lambda}
\biggl[\Bigl(a e^{-i\phi}-a^{\dagger} e^{i\phi}\Bigr)\cos\theta
				{\partial\over\partial\theta}
-{i\over\sin\theta}\Bigl(a e^{-i\phi}+a^{\dagger} e^{i\phi}\Bigr)
				{\partial\over\partial\phi}\biggr]
\\
\nonumber
&=&{1\over \sqrt2 p\Lambda}
\biggl[\Bigl(b e^{i\phi}- b^{\dagger} e^{-i\phi}\Bigr)\cos\theta
				{\partial\over\partial\theta}
+{i\over\sin\theta}\Bigl(b e^{i\phi}+ b^{\dagger} e^{-i\phi}\Bigr)
				{\partial\over\partial\phi}\biggr].
\end{eqnarray}

\section{Frequency Sums}
\label{sec:sums}

The sum of the values of a response
function at
Matsubara frequencies $i\omega_n=(2n +1)\pi i T$
in the upper half plane can be written as an integral
\begin{equation}
T\sum_{n=0}^\infty K(i\omega_n)={1\over 4\pi i}\int_{-\infty}^{+\infty}
 \tanh 
		\Bigl({\omega\over 2T}\Bigr) K(\omega).
\label{sum1}
\end{equation}
If the response function varies
slowly over the scale $\omega\sim T$, and
the tangent can be replaced with a
step function, so that
\begin{equation}
T\sum_{n=0}^\infty K(i\omega_n)\approx
{1\over 4\pi i}\Bigl( \lim_{\omega\rightarrow\infty} F(\omega) + 
\lim_{\omega\rightarrow-\infty}F(\omega) - 2 F(0)\Bigr),
\label{sum}
\end{equation} 
where $ K(\omega)  = (d F(\omega)/d\omega)$.

First consider the sum that appears 
in the quasiclassical contribution
to the longitudinal current
\begin{equation}
S=
\sum_{\omega_n}{\Bigl(g-g(-)\Bigr)\over i\widetilde\omega_0}.
\end{equation}
Since the frequency $\widetilde\omega_0$ can be replaced with bare
frequency in the outer frequency range, but is renormalized in the 
intermediate range, we consider the sum separately in the two regions.
In the outside region, transforming the sum in the lower half plane into
a sum over the frequencies in the upper half plane,
\begin{equation}
S_{out}=-{2\over\omega_0}\sum_{\omega_n>0}\Bigl(P(+)-P\Bigr)
=-2i\sum_{\omega_n>0}\Bigl({\partial P \over \partial \bar\omega}+
			{\bar\omega\over 2}
{\partial^2 P\over \partial\omega^2}\Bigr).
\end{equation}
after analytic continuation and expansion in $\bar\omega$.
Using Eq. (\ref{sum})
we obtain
\begin{equation}
\label{qpout1}
S_{out}={1\over\pi}\Bigl[ (P-1) + 
\Bigl({\Lambda\bar\omega\over v\sin\theta}\Bigr)
 	{\partial P\over\partial\omega}\Bigr],
\end{equation}
where the values of the functions are computed at $\omega=0$.
To evaluate the sum in the intermediate frequency range
to leading functional order in $\Delta^2$, we write
\begin{equation}
\label{sintqp}
S_{int}=-i\sum_{\omega_n>0}^{\omega_0}{P+P(-)\over i\omega_0+ (i/2\tau)
			(\langle P\rangle+\langle P(-)\rangle )}
=-2i\sum_{\omega_n>0}^{\omega_0}P{1\over i\omega_0+ (i/2\tau)
			(\langle P\rangle+\langle P(-)\rangle )}.
\end{equation}
Adding and subtracting
 the contribution of  a normal metal, so that the
remaining sums are convergent at high frequency, 
we obtain after analytic
continuation
\begin{equation}
\label{qpin}
S_{int}={i\bar\omega\tau\over\pi}\Biggl[\Bigl(P-1\Bigr)
+\biggl[1+ \langle \Bigl(1-P\Bigr)\rangle\biggr]\Biggr],
\end{equation}
and
\begin{equation}
\label{qpsum}
S={1\over\pi}
\Biggl[(P-1) +i\bar\omega\tau\Bigl[P + \langle (1-P)\rangle
-\sqrt\pi W''\Bigl({i\Lambda\over l\sin\theta}\Bigr)
\Bigl({2\Lambda\Delta\over v\sin\theta}\Bigr)^2
\Bigl({\Lambda\over l\sin\theta}\Bigr)P^3\Bigr]\Biggr].
\end{equation}

The vertex appearing in the calculation of the
dynamical fluctuations of the order parameter is
proportional to
\begin{equation}
\label{sumw1}
\sum_{\omega_n} {gW'-g(-)W'(-)\over i \widetilde\omega_0}.
\end{equation}
Since the amplitude of the dynamical fluctuations
only has to be evaluated to zeroth order in the superconducting order
parameter, it is sufficient here to replace the Green's function 
in the renormalized frequency by its normal state value.
Since this sum is well-behaved at high frequency, we easily obtain
\begin{equation}
V=\sum_{\omega_n} {gW'-g(-)W'(-)\over i \widetilde\omega_0}=
-2i\Bigl({1\over i \widetilde\omega_0 +i/\tau }-{1\over i\widetilde\omega_0}\Bigr)
\sum_{\omega_n>0}\Bigl(PW'-P(+)W'(+)\Bigr).
\end{equation}
After analytic continuation we find
\begin{eqnarray}
V&=&{-2i\over 1 -i \bar\omega\tau}
{2\Lambda\over v\sin\theta}
\sum_{\omega_n>0}\Biggl\{ W''(u_n)+\Bigl({\Lambda\bar\omega\over v\sin\theta}\Bigr)
				W^{(3)}(u_n)\Biggr\}
\\
\nonumber
&=&{1\over \pi}{1\over 1-i\bar\omega\tau}
\Biggl\{W'\Bigl({i\Lambda\over l\sin\theta}\Bigr)
	 +{\Lambda\bar\omega\over v\sin\theta}
				 W''\Bigl({i\Lambda\over l\sin\theta}\Bigr)
					\Biggr\}.
\end{eqnarray}

The sum in the fluctuation propagator in Eq. (\ref{C}) is 
easily evaluated in a similar fashion
\begin{eqnarray}
&&\sum_{\omega_n} \Bigl[ig{\rm sgn }(\omega_n) W(u_n)-{i\over 2}
             \Bigl(g+g(-)\Bigr){\rm sgn}(\omega_n)
			\Bigl(W(U_n)+{1\over 2}W''(U_n)\Bigr)\Bigr]
\\
\nonumber
&&
= \sum_{\omega_n>0}\Bigl[2 PW -\Bigl(P+P(+)\Bigr)
			\Bigl(W(U_n^+)+{1\over 2}W''(U_n^+)\Bigr)\Bigr]
+{1\over 2} \sum_{\omega_n>0}^{\omega_0}
\Bigl(P-P(-)\Bigr)\Bigl(W(U_n)+{1\over 2}W''(U_n)\Bigr){\rm sgn}(\Omega_n),
\end{eqnarray}
where $2U_n^+=u_n+u_n^+$.
Since \cite{Abramowitz}
\begin{equation}
W(U_n)+{1\over 2}W''(U_n)=-U_n W'(U_n),
\end{equation}
the contribution of the last term to the final results is at least
of order $\bar\omega^3$. The remaining terms give, after expansion in
the external frequency,
\begin{eqnarray}
\label{sumden11}
&&T\sum_{\omega_n} \Bigl[ig{{\rm sgn} }(\omega_n) W(u_n)-{i\over 2}
			\Bigl(g+g(-)\Bigr){\rm sgn}(\Omega_n)
			\Bigl(W(U_n)+{1\over 2}W''(U_n)\Bigr)\Bigr]
\\
\nonumber
&&
= -T\sum_{\omega_n>0}\Bigl[W''(u_n)+ 
\Bigl({2\Lambda\bar\omega\over v\sin\theta}\Bigr)
		\Bigl(W'(u_n)+{1\over 2}W^{(3)}(u_n)\Bigr)\Bigr]
\\
\nonumber
&&
= {1\over 2\pi i}{v\sin\theta\over 2\Lambda}\Bigl\{
		W'\Bigl({i\Lambda\over l\sin\theta}\Bigr)+
		\Bigl({2\Lambda\bar\omega\over v\sin\theta}\Bigr)
               \Bigl[W\Bigl({i\Lambda\over l\sin\theta}\Bigr)
			+{1\over 2}W''\Bigl({i\Lambda\over l\sin\theta}\Bigr)
			\Bigr]		\Bigr\}.
\end{eqnarray}

\section{Fluctuations of the order parameter.}
\label{sec:C}

Our starting point here is Eq. (\ref{f1smod}) for the linear correction
to the anomalous propagator
\begin{eqnarray}
\label{f1C}
f_1&=&{e {\bf v A}(f-f(-)) \over i\widetilde\omega_0} 
	+ i(g+g(-))
\Bigl[ 2 \widetilde \Omega_n + {\bf v} (\nabla - 2 i e {\cal A})\Bigr]^{-1}
\Delta_1
\\
\nonumber
&& +i (2\tau)^{-1}(g+g(-))
\Bigl[ 2 \widetilde \Omega_n + {\bf v} (\nabla - 2 i e {\cal A})\Bigr]^{-1}
\langle f_1 \rangle
\\
\nonumber
&&
+ i (2\tau)^{-1}{e{\bf v A} (g-g(-)) \over i\widetilde\omega_0}
\Bigl[ 2 \widetilde \Omega_n + {\bf v} (\nabla - 2 i e {\cal A})\Bigr]^{-1}
\Bigl(\langle f \rangle + \langle f(-)\rangle\Bigl).
\end{eqnarray}
If we define the angular average of $f_1$ by
\begin{equation}
\langle f\rangle=S(\omega_n)|1\rangle,
\end{equation}
we find, after carrying out the angular integration and ignoring terms of
order $\Lambda/l$
\begin{equation}
\label{Ssimple}
S=
\sqrt\pi {\Lambda\over v}
\int_0^\pi {d\theta  \over 4}
\Biggl\{ 2iev A \sin\theta \Big(-{i\over \sqrt 2}\Bigr)
{gW'-g(-)W'(-) \over i\widetilde\omega_0} 
+i(g+g(-)) 2 C
{\rm sgn}(\Omega_n)\Bigl(W(U_n)+{1\over 2}W''(U_n)\Bigr)\Biggr\}.
\end{equation}
We now use Eq. (\ref{helpC}) to determine the 
amplitude of the excited mode of the order parameter
$C=\pi{\rm g} N(0) \sum_n S$
\begin{eqnarray}
\label{c}
&&C\Bigl[1-\pi{\rm g} N(0)\sqrt\pi {\Lambda\over v}
 \sum_n \int_0^\pi d\theta {i\over 2}(g+g(-))
{\rm sgn}(\Omega_n)\Bigl(W(U_n)+{1\over 2}W''(U_n)\Bigr)\Bigr]
\\
\nonumber
&&=\pi{\rm g} N(0)\sqrt\pi {\Lambda\over v}
\sum_n \int_0^\pi{\sqrt2\over 4}ev A \sin\theta d\theta
{gW'-g(-)W'(-) \over i\widetilde\omega_0}.
\end{eqnarray}
We can now use the gap equation to eliminate the need for a frequency cutoff
and obtain
\begin{eqnarray}
\label{C1}
&&C\sum_n \int_0^\pi d\theta 
\Bigl[ig{\rm sgn}(\omega_n)W(u_n)-{i\over 2}(g+g(-))
{\rm sgn}(\Omega_n)\Bigl(W(U_n)+{1\over 2}W''(U_n)\Bigr)\Bigr]
\\
\nonumber
&&\qquad=\sum_n \int_0^\pi{\sqrt2\over 4}ev A \sin\theta d\theta
{gW'-g(-)W'(-) \over i\widetilde\omega_0}.
\end{eqnarray} 
It follows that to leading order
\begin{equation}
\label{Cvalue1}
C={ie\Lambda A \sqrt2\over 1-i\bar\omega\tau}.
\end{equation}
We also give the expression for the
distribution function $f_1$. Neglecting the contributions of
order $(\Lambda/l)$, we use the Eq. (\ref{oper1}) to compute
\begin{eqnarray}
\label{f1value}
&&f_1=e{\bf v A}{f-f(-)\over i\widetilde\omega_0}+
	i\Bigl(g+g(-)\Bigr)
\Bigl[2\widetilde\Omega_n + {\bf v}\bigl(\nabla-2ie{\cal A}\bigr)\Bigr]^{-1}
\Delta_1
\\
\nonumber
&&={2i\over i\widetilde\omega_0}\sqrt\pi
e\Lambda A\cos\phi
\sum_{m=0}^\infty {1\over\sqrt{m!}}e^{im\phi}\Bigl(-{i\over\sqrt2}\Bigr)^m
\times
\biggl\{ gW^{(m)}{\rm sgn}^{m+1}(\omega_n)-
g(-)W^{(m)}(-){\rm sgn}^{m+1}(\omega_n-)
\biggr\} |m\rangle
\\
\nonumber
&&+i\Bigl(g+g(-)\Bigr)C{\sqrt\pi\Lambda\over v\sin\theta}
\sum_{m=0}^\infty{1\over\sqrt{m!}}\Bigl(-{i\over\sqrt2}\Bigr)^m
{\rm sgn}^{m+1}(\Omega_n)
\\
\nonumber
&&\times
\biggl\{\sqrt{m+1}e^{im\phi}W^{(m)}(U_n)|m+1\rangle
+{i\over\sqrt2}{\rm sgn}(\Omega_n)e^{i(m-1)\phi}W^{(m+1)}(U_n)|m\rangle
\biggr\}.
\end{eqnarray}
The evaluation of $\bar C$ is analogous to the calculation given above.

In the transverse response calculation our
starting point here is Eq. (\ref{fhall}) 
for the linear, in cyclotron frequency, correction
to the anomalous propagator
\begin{equation}
\label{df1app}
\delta f=
\bigl[ 2 \widetilde \Omega_n + {\bf v}_f (\nabla - 2 i e {\cal A})\bigr]^{-1}
\biggl\{-{e\over 2m}{\bf A}\Bigl(\nabla -2ie{\cal A}\Bigr)\Bigl(f-f(-)\Bigr)
+i\Delta \delta g_e + i\delta\Delta_1\Bigl(g+g(-)\Bigr)\biggr\}.
\end{equation}
The solution of this equation follows exactly the steps
described in the previous section. First we solve for
the coefficient $\delta C$ in $\delta\Delta_1=\delta C|1\rangle$.
As $\delta C=\sum_n\int d^2 s \delta f$,
the denominator of the expression for $\delta C$
is the propagator for the first excited mode of the order parameter,
as it was for $C$. To evaluate contribution of each of the
driving terms we notice that for our choice of {\bf A}
\begin{equation}
{\bf A}\Bigl(\nabla -2ie{\cal A}\Bigr)={\bf A}\nabla_x=
A{1\over \sqrt 2 \Lambda}\Bigl[ a-a^{\dagger}\Bigr].
\end{equation}
Explicit evaluation of this term
using  the expression for the function $f$ from Eq. (\ref{f})
shows that it contributes at order $\Lambda/l$ compared to leading order
terms. The driving term  $i\Delta \delta g_e$ only 
contributes in the intermediate frequency range since
\begin{equation}
\delta g_e= -2i \omega_c e v A \sin\theta\sin\phi 
{g-g(-)\over(i\widetilde\omega_0)^2},
\end{equation}
and, to the order
in which we work, $g-g(-)$ vanishes in the outer range.
Then in analogy to the solution outlined above we
obtain
\begin{equation}
\delta C_1= \delta\bar C_1=ie\Lambda A \sqrt2 (\bar\omega\tau)(\omega_c\tau).
\label{dC1}
\end{equation}
and, for the angular average of the function $\delta f$
needed to calculate the Thompson contribution to the conductivity,
\begin{eqnarray}
\label{thomhall2}
\langle \delta f_1 \rangle=
\sqrt{2\pi} e\Lambda A {\omega_c\over (i\widetilde\omega_0)^2}
\int_0^\pi {d\theta  \over 4}\biggl\{ \sin\theta \Bigl( g-g(-)\Bigr)
W'(U_n)
+2i\delta C \Bigl( g+g(-)\Bigr) {\rm sgn}(\Omega_n)
	\Bigl(W(U_n)+{1\over 2} W''(U_n)\Bigr)\biggr\} |1\rangle
\\
\langle \delta f_1^{\dagger} \rangle=
\sqrt{2\pi} e\Lambda A {\omega_c\over (i\widetilde\omega_0)^2}
\int_0^\pi {d\theta  \over 4}\biggl\{ \sin\theta \Bigl( g-g(-)\Bigr)
W'(U_n)
+2i\delta \bar C \Bigl( g+g(-)\Bigr) {\rm sgn}(\Omega_n)
	\Bigl(W(U_n)+{1\over 2} W''(U_n)\Bigr)\biggr\} \langle 1|.
\end{eqnarray} 
In the intermediate region $g+g(-)=0$, and only the first term
in each function contributes to the conductivity to leading order
in $(\Lambda/l)$.
%%%%%%%%%%%%%%%%%%%%%%%%%%%%%%%%%%%%%%%
 
\newpage
\begin{figure}
%\epsfysize=4.25in
%\center{\hspace 3 \epsfbox{fig1.ps}}
\caption{Longitudinal resistivity as a function of the reduced magnetic field.}
\label{fig:rxx}
\end{figure}
\begin{figure}
%\epsfysize=4.25in
%\center{\hspace 3 \epsfbox{fig2.ps}}
\caption{Transverse conductivity as a function of the reduced magnetic field.}
\label{fig:sxy}
\end{figure}
\begin{figure}
%\epsfysize=4.25in
%\center{\hspace 3  \epsfbox{fig3.ps}}
\caption{Hall angle as a function of the reduced magnetic field.}
\label{fig:tan}
\end{figure}

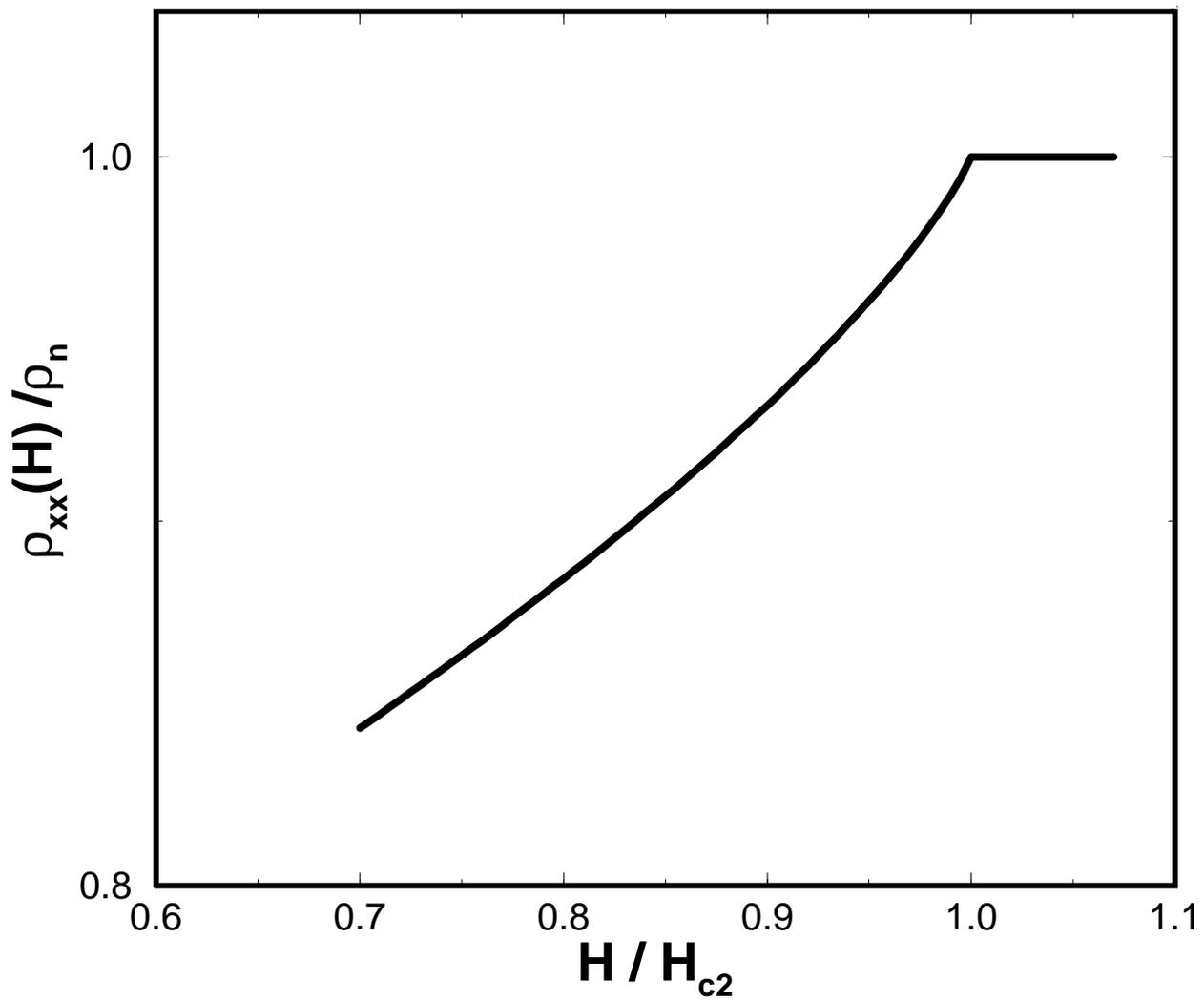
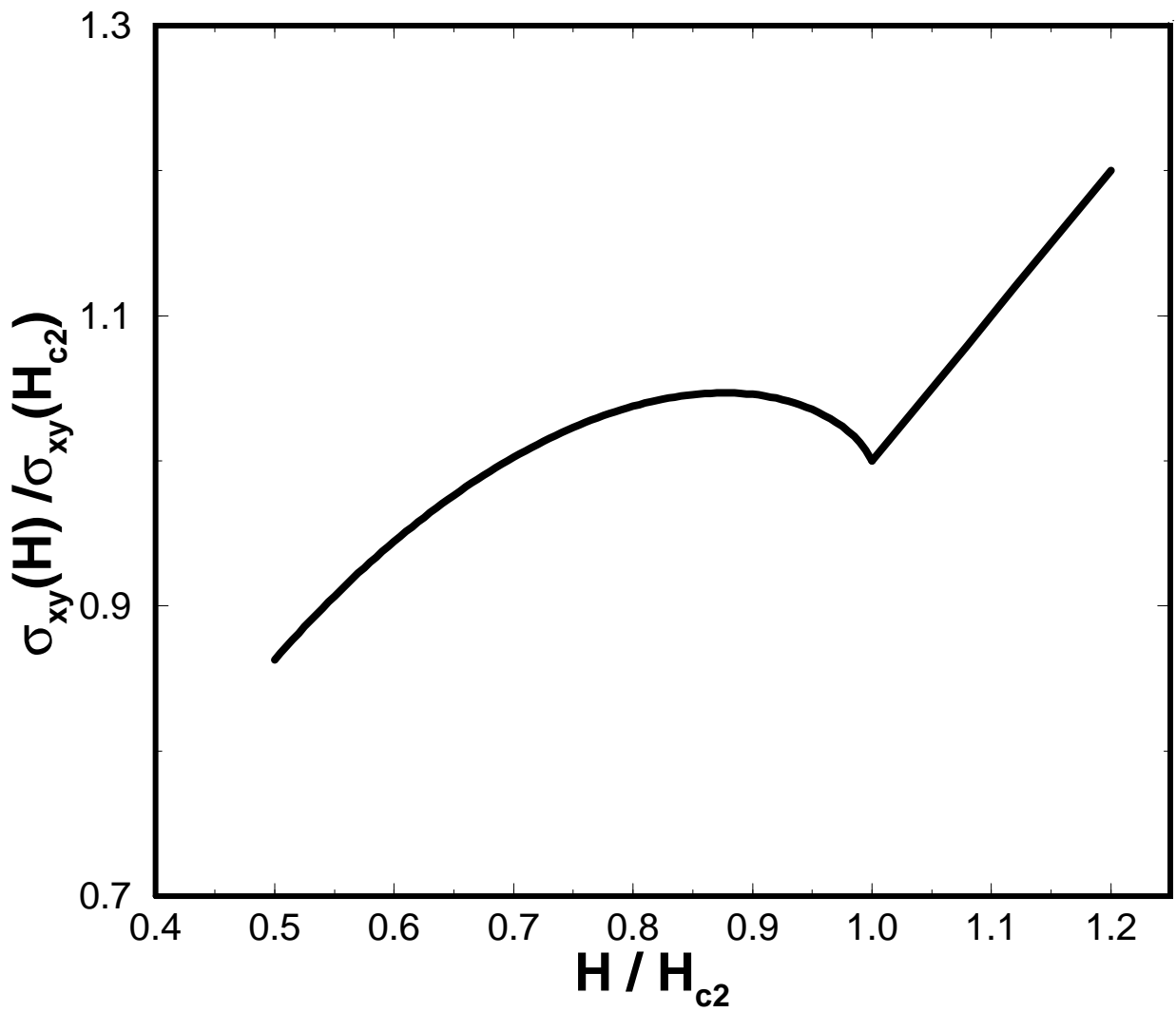
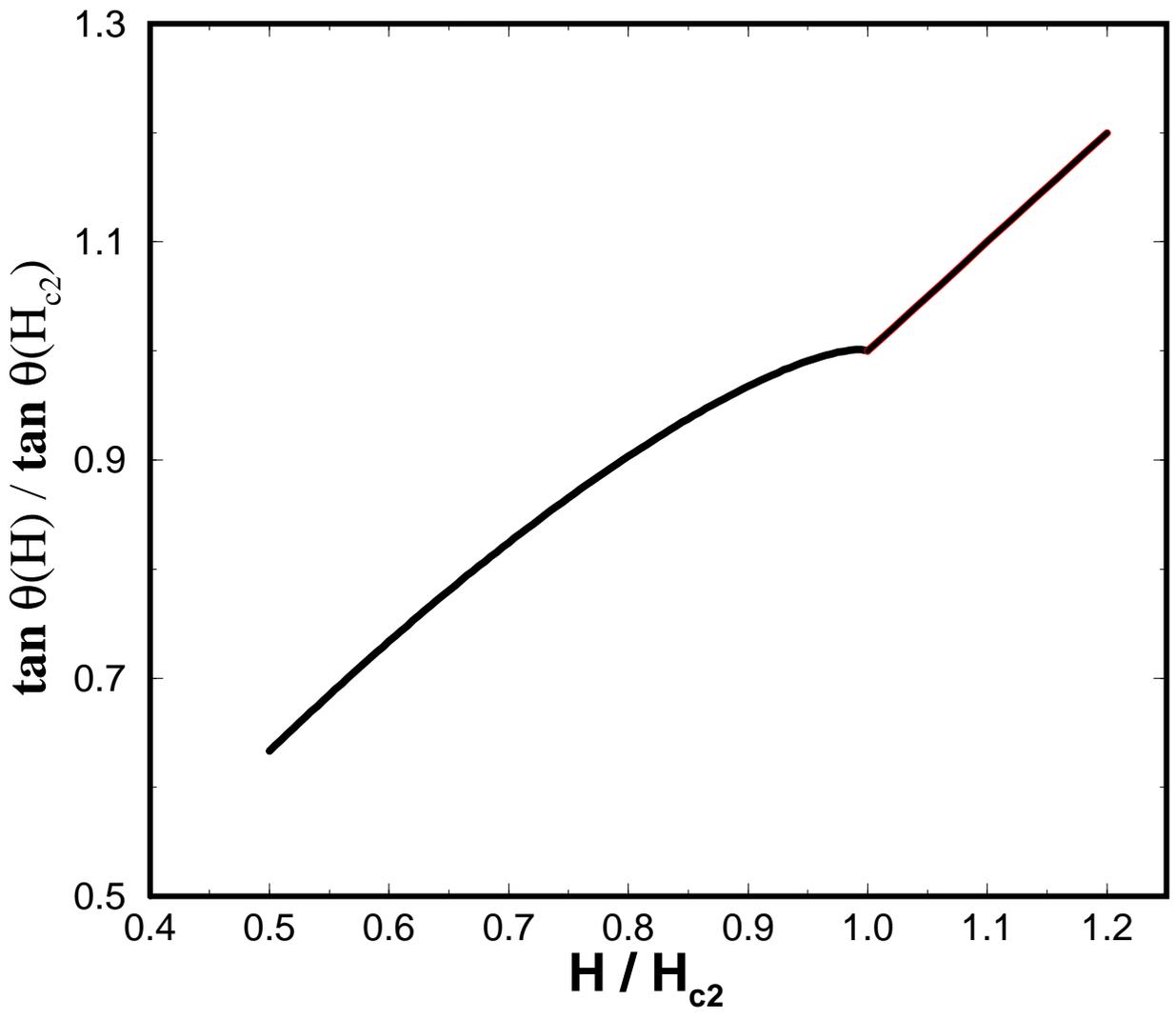
\end{document}